\documentclass[sn-mathphys-num, iicol]{sn-jnl}% [iicol]
\usepackage{graphicx}%
\usepackage{multirow}%
\usepackage{amsmath,amssymb,amsfonts}%
\usepackage{amsthm}%
\usepackage{mathrsfs}%
\usepackage[title]{appendix}%
\usepackage{xcolor}%
\usepackage{textcomp}%
\usepackage{manyfoot}%
\usepackage{booktabs}%
\usepackage{algorithm}%
\usepackage{algorithmicx}%
\usepackage{algpseudocode}%
\usepackage{listings}%
\usepackage{xcolor}
\usepackage{multirow}
\usepackage{amsbsy,amsfonts,amsfonts,amssymb,amsmath,epsfig}
\usepackage{graphicx}
\usepackage{graphics}

\usepackage{url}

\usepackage{natbib}
\usepackage{wrapfig}

\usepackage{color, colortbl}
\definecolor{Gray}{gray}{0.9}
\definecolor{LightCyan}{rgb}{0.88,1,1}
\usepackage[first=0,last=9]{lcg}

\usepackage{color, colortbl}
\definecolor{Gray}{gray}{0.9}
\definecolor{LightCyan}{rgb}{0.88,1,1}
\usepackage[first=0,last=9]{lcg}

\usepackage{color}

\usepackage{subcaption}
\theoremstyle{thmstyleone}%
\theoremstyle{thmstyletwo}%

\theoremstyle{thmstylethree}%

\begin{document}

\title[Article Title]{A New Wireless Image Transmission System Using Code Index Modulation and Image Enhancement for High-Rate Next Generation Networks}


\author[1,2]{\fnm{Burak Ahmet } \sur{Ozden}}\email{bozden@yildiz.edu.tr}

\author*[1]{\fnm{Erdogan} \sur{Aydin}}\email{erdogan.aydin@medeniyet.edu.tr}

\author[2]{\fnm{Ahmet} \sur{Elbir}}\email{aelbir@yildiz.edu.tr}

\author[1]{\fnm{Filiz} \sur{Gurkan}}\email{filiz.gurkan@medeniyet.edu.tr}

\affil*[1]{\orgdiv{Department of Electrical and Electronics Engineering}, \orgname{Istanbul Medeniyet University},\orgaddress{ \city{Uskudar}, \postcode{34857}, \state{Istanbul}, \country{Turkey}}}

\affil[2]{\orgdiv{Department of Computer Engineering}, \orgname{Yildiz Technical University}, \orgaddress{ \city{Davutpasa}, \postcode{34220}, \state{Istanbul}, \country{Turkey}}}


\abstract{With the development of wireless network technologies, the wireless image transmission area has become prominent. The need for high resolution, data traffic density, widespread use of multimedia applications, and the importance of high rate and reliable image transmission in medical and military fields necessitate the design of novel and high-performance wireless image transmission systems. This paper proposes a code index modulation (CIM)-based image transmission (CIM-IT) system that utilizes spreading code index and quadrature amplitude modulation (QAM) symbol for image transmission over a wireless channel. The proposed CIM-IT system maps bits to each pixel value of the image to be transmitted and transmits these bits over a wireless channel using a single-input and multiple-output system comprising code index modulation and QAM techniques. At the receiver, the active spreading code index and the selected QAM symbol are estimated using a despreading-based maximum likelihood detector, and the corresponding bits are obtained. The image conveyed from the transmitter is then reconstructed at the receiver side using the pixel values corresponding to the bits. The obtained noisy image is enhanced using important enhancement filters. In addition, an advanced filter is proposed to improve the transmitted degraded image with optimum results. Furthermore, error performance, spectral efficiency, energy efficiency, and throughputof the CIM-IT system are performed and the results are compared with traditional wireless communication techniques. 

%According to the results obtained, it is shown that the CIM-IT system provides better results than traditional communication systems in every respect.
}




\maketitle

\section{Introduction}\label{sec1}
Nowadays, challenges such as the widespread use of mobile devices, the increase of remotely controllable devices, and high data transfer traffic have emerged. In addition, due to the widespread use of mobile devices, wireless communication systems have become dependent on wireless networks and as a result, high data rate, low energy consumption, high reliability, and high spectral efficiency are required. 6G technology stands out as the next-generation wireless network that aims to meet many predicted needs. 6G is predicted to provide a higher speed, lower latency, and higher bandwidth than 5G. Also, 6G technology makes it possible not only for communication between people but also for communication between machines, which is known as the Internet of Things \cite{6GTATARIA, 6GBARIAH, 6GGUO, RISAS}. 
%However, meeting the above-mentioned needs and laying the foundation for 6G technology is only possible by designing new high-performance and high-data-rate wireless communication systems. 

Multiple-input multiple-output (MIMO) systems are one of the pioneering technologies in wireless communication technologies to make wireless data transfer more efficient. 
%MIMO systems utilize multiple antennas to improve spectral efficiency, error performance, and channel capacity. 
The advantages of MIMO systems include increasing data rates, increasing energy efficiency, improving link quality, and providing more secure wireless communications. However, MIMO systems also face some challenges. Challenges such as data loss, inter-antenna synchronization problems, inter-channel interference, latency, and environmental factors can negatively affect the performance of MIMO systems \cite{deepmimo, erdoganSM, STBCMIMO, fatihanten}. These challenges can be largely overcome when MIMO systems are augmented with index modulation (IM) that offers a new and efficient approach to wireless data transmission. In addition to the information traditionally transmitted in the symbol, index modulation offers a highly efficient wireless data transmission approach by using the active and passive states of the parameters used in wireless data transmission as a modulation space. IM schemes are innovative transmission techniques that exploit various transmission entities, including antenna indices, subcarrier indices, spreading code indices, or time-slot indices, to transmit extra information bits alongside conventional $M$-ary symbols \cite{IMSURVEY1, IMSURVEY2, IMSURVEY3, burakris}. IM techniques offer many substantial advantages such as increased spectral efficiency, improved error performance, enhanced security, lower power consumption, and reduced complexity for next-generation wireless networks \cite{sugiroindex, ertoindex, whenindex, codeozden}.

\begin{figure*}[t]
\centering{\includegraphics[width=0.9\textwidth]{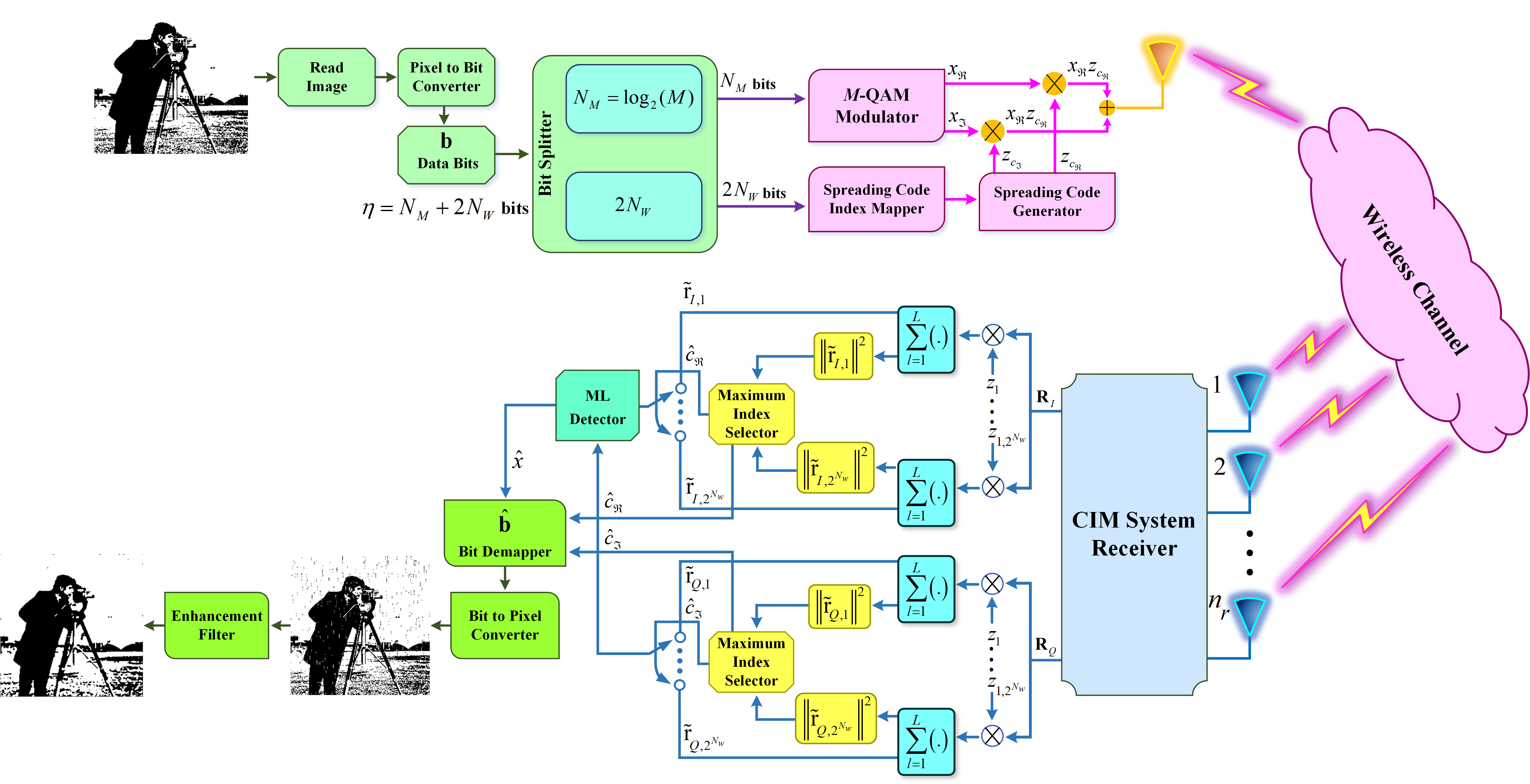}}	\caption{System model of the CIM-based image transmission system.}
	\label{system_model} 	
\end{figure*}

Multiple access systems are technologies that enable multiple users to access a communication network simultaneously. Code division multiple access (CDMA) is an important technique among multiple access techniques that enables the most effective sharing of the spectrum. CDMA systems provide many advantages such as serving a large number of users, enhanced security, high spectral efficiency, high flexibility, resilience to disturbance, improved quality of service, efficient use of bandwidth, and increased cellular capacity, by using unique codes \cite{CDMA1, CDMA2, CDMA3, CDMA4, CDMA5}. Code index modulation (CIM) is a type of CDMA technique that uses orthogonal spreading codes to transmit additional information in addition to the information carried in the symbol \cite{Cogen2018siu, burakwcnc}. Thanks to the use of orthogonal spreading codes in the CIM technique, the transmitted signal is affected less by disruptive factors such as noise, fading, and interference. The CIM technique enables secure data transmission and ensures low error rates even at low power levels and low signal-to-noise ratio (SNR) values. CIM has been proposed for direct sequence spread spectrum (DS-SS) communication and has been shown to improve energy efficiency, reliability of data transmission, and spectral efficiency. CIM has also been used in 6G downlink transmission via rate splitting space division multiple access based on code index modulation \cite{6gcim, kadoumcim, fatihris}. The authors propose a new wireless communication system called CIM-SMBM that combines index modulation-based techniques to meet the needs of future wireless networks in \cite{burakcim}. Analyses of the proposed system's bit error rate, throughput, energy efficiency, and complexity over Rayleigh fading channels are presented. In \cite{CIMGCSDCSK}, a new scheme is proposed that combines the CIM and the general code-shifted differential chaos shift keying (GCS-DCSK) system, called CIM-CS-DCSK, to carry additional bits using specific indices selected by an integer sequence matching method, achieving higher spectrum efficiency and data rate than GCS-DCSK. A new communication scheme, which uses CIM and quadrature spatial modulation (QSM), called CIM-QSM, is presented to achieve high spectrum and energy efficiency in MIMO systems by transmitting information through modulated symbols, activated antenna indices, and spreading code indices in \cite{aydinQSM}.

Nowadays, secure and high-rate wireless image transmission has become an essential need. Wireless image transmission is used in many fields such as medicine, security, industrial control, military, and entertainment. However, wireless image transmission has many challenges including low data rates, transmission error, noise, high latency, and insufficient bandwidth \cite{imagesurvey1, imagesurvey2}. Therefore, researchers in this field have proposed many new approaches and systems to develop and improve wireless image transmission technologies. \cite{haciimage} proposes a novel method for cooperative image transmission in wireless sensor networks, focusing on efficient relayed image transmission through wireless channels with optimum image quality and bit error rate performances, utilizing lightweight image quality improvement, and the decode-forward method at relay nodes. In \cite{zhengimage}, a millimeter wave wireless Hadamard image transmission technique using cyclic Hadamard transform for compression, sub-sampling, and reconstruction is proposed in the study, which is compatible with MIMO and can be implemented in software or hardware and has been shown to have a significant increase in image transmission speed, with potential for even faster transmission in the upcoming 6G wireless network. Also, numerous studies have been presented in the literature regarding wireless image transmission \cite{imaiterat1, imaiterat2, imaiterat3, imaiterat4, imaiterat5, sonimage1, sonimage2}.

This paper proposes a wireless communication system for high-performance wireless image transmission by using the CIM technique, an IM scheme with many advantages, in a single-input and multiple-output (SIMO) structure. The proposed CIM-based image transmission system is called CIM-IT for short. The main contributions of this article can be summarized as follows:

\begin{enumerate}

\item The proposed CIM-IT system provides wireless image transmission with lower bit error than conventional wireless communication systems.
\item The throughput, energy efficiency, and spectral efficiency analysis of the CIM-IT system and traditional quadrature amplitude modulation (QAM) and quadrate phase shift keying (QPSK) systems are derived. 
\item The proposed CIM-IT system is compared with traditional systems such as QAM and QPSK and it is observed that for the same spectral efficiency, the proposed system has a better BER performance.
\item The performance of the CIM-IT system for wireless image transmission for different system parameters is investigated. In this way, it is shown which system parameter affects the performance and how.
\item Significant enhancement filters are applied to the image at the receiver, which is distorted by the fading effect of the wireless channel and noise, to produce an enhanced image at the receiver.

\end{enumerate}

\section{CIM-based Image Transmission  System Model}
The CIM-IT system uses spreading code indices and QAM modulation for high-data-rate wireless data transmission. The CIM-IT system model, depicted in Fig. \ref{system_model}, utilizes $M$-ary quadrature amplitude modulation ($M$-QAM), $n_r$ receive antennas, one transmit antenna, and DS-SS with $2^{N_W}$ orthogonal spreading Walsh–Hadamard codes $ \{\textbf{z}_1,  \ldots, \textbf{z}_{N_W}\}$ at the transmitter. Each spreading code consists of $L$ chips (with a chip period of $T_z$) and can be represented as a vector, such as $\textbf{z}_c = [z_{c,1},  \ldots, z_{c, L}]^T$. 

In the first step of the process, the gray image is accessed and read, as depicted in Fig. \ref{system_model}. This is achieved by utilizing an appropriate method to extract the image data. Subsequently, the transformation from pixel to binary representation takes place for each pixel of the read image. This conversion process, which is performed, assigns a binary value to each pixel. As a result, a set of data bits represented by $\textbf{b}$ is obtained, representing the binary representation of the image. The data bits $\textbf{b}$ are aimed to be transmitted over a wireless Rayleigh fading channel with the help of some modulation techniques.

Considering the transmitter structure of the CIM-IT system, the bit sequence $\textbf{b}$ of size $\eta\times1$ is applied to the bit splitter. Bit splitter divides the $\eta$ bits into two subgroup blocks. The first subgroup, with a length of $N_M=\mathrm{log}_2 (M)$ bits is mapped to the $M$-QAM symbol $ x=x_{\Re} + jx_{\Im}$ while the second subgroup, with a length of $2N_W$ bits, is mapped to the two spreading code indices. Here, according to the data bits, the symbol to be selected from the $M$-QAM symbol space and the spreading codes to be selected from the code set are decided. As a result, the total bits transmitted during a symbol period in a CIM system, which is defined as the spectral efficiency, is expressed as 
$\eta=2N_W+N_M$.
%follows:
%\begin{eqnarray}\label{spectralef}
%	\eta=2N_W+N_M.
%\end{eqnarray}
Then, the spreading codes ($\textbf{z}_{\Re}$ and $\textbf{z}_\Im$) generated from the spreading code generator are then multiplied by the real ($x_{\Re}$) and imaginary  ($x_{\Im}$) parts of the symbol $x$ generated from the $M$-QAM modulator. Finally, from $I$ and $Q$ components, the symbol multiplied by the spreading codes to be transmitted over the Rayleigh fading wireless channel is transmitted to the receiver via an antenna placed at the transmitter.

Considering the receiver structure of the CIM-IT system, the signal at the receiver, subject to fading and noise effects, is expressed in baseband terms as follows:
\begin{eqnarray}\label{baseband}
r(t)  &  = & \sum_{l=1}^{L} \bigg[x_\Re\,z_{c_{\Re},l}\,u\big(t-lT_c\big)\cos(2\pi f_ct) + x_\Im\,z_{c_{\Im},l} \nonumber \\
&  \times & u\big(t-kT_c\big)\sin(2\pi f_ct) \bigg]h(t)+w(t), 
\end{eqnarray}
where, the signal is shaped by a unit rectangular pulse shaping filter, denoted as $u(t)$, over the time period from 0 to $T_z$. The channel impulse response, $h(t)$ corresponds to the Rayleigh fading channel. Additionally, the term $w(t)$ represents a complex Gaussian random process with a zero mean and a variance of $N_0$.

The $l\text{th}$ noisy chip signal for both the $I$ and $Q$ components is obtained after perfect carrier detection and sampling at the output of the channel. Hence, the received signal in (\ref{baseband}) can be re-expressed by separating it into $I$ and $Q$ components as follows:
\begin{eqnarray}\label{receiveIQ1}
	 r_{I} &   = & x_\Re\,z_{c_{\Re}}\, h  + w_{I}, \: \: \: r_{Q}      =   x_\Im\,z_{c_{\Im}}\, h  + w_{Q},    
\end{eqnarray}
where, $c_\Re,c_\Im=1,2,\ldots,2^{N_W}$ and $h$ represents the channel coefficient for a Rayleigh fading channel, which has zero mean and unit variance. $w_I$ and $w_Q$ symbolize additive white Gaussian noise (AWGN) expressions with zero mean and $N_0$ variance for the $I$ and $Q$ components, respectively. In vector-matrix form, the expression in (\ref{receiveIQ1}) can be rewritten as follows:
\begin{eqnarray}\label{eq4}
\textbf{R}_I  &  = & x_\Re\,\textbf{h}\, \textbf{z}^T_{c_{\Re}}  + \textbf{W}_I  \nonumber \\
\textbf{R}_Q  & =& x_\Im\,\textbf{h}\, \textbf{z}^T_{c_{\Im}}  + \textbf{W}_Q,
\end{eqnarray}
where, the channel vector $\textbf{h}$ has dimensions $n_r \times 1$ and $\textbf{h} = \big[h_1, h_2, \ldots, h_{n_r}\big]^T$. $\textbf{z}_{c_{\Re}}$ and $\textbf{z}_{c_{\Im}}$, both sized $L \times 1$, are spreading codes. Also, $\textbf{W}_I$ and $\textbf{W}_Q$ are AWGN matrices. 

As shown in Fig. 1, the despreaded output of the $c\text{th}$ correlator for both the $I$ and $Q$ components is obtained by multiplying the $\textbf{R}_I$ and $\textbf{R}_Q$ with their corresponding spreading codes $\textbf{z}_c$ in each branch and summing them over the period $T_x = LT_z$. Consequently, the despreaded output for both the $I$ and $Q$ components at the $c\text{th}$ correlator can be described as follows:
\begin{eqnarray}\label{corelator1}
\tilde{\textbf{r}}_{I,c}  &  = & \textbf{R}_I \, \textbf{z}_c=\Big[\mathbf{r}_{I,1}\, \mathbf{r}_{I,2}\, \ldots \,\mathbf{r}_{I,L}\Big]\textbf{z}_c \nonumber \\
\tilde{\textbf{r}}_{Q,c}  &  = & \textbf{R}_Q \, \textbf{z}_c=\Big[\mathbf{r}_{Q,1}\, \mathbf{r}_{Q,1}\, \ldots \, \mathbf{r}_{I,L}\Big]\textbf{z}_c,
\end{eqnarray}
where, the $k\text{th}$ elements of $\tilde{\textbf{r}}_{I,c} $and $\tilde{\textbf{r}}_{Q,c}$ can be defined as follows:

\begin{eqnarray}\label{corelator2}
\tilde{r}^k_{I,c} &  = & \sum_{l=1}^{L} z_{c,l}\,r_{I,l} = \sum_{l=1}^{L} z_{c,l}\,\Big(x_{\Re}\,z_{c_{\Re},l}\, h  + w_{I,l}\Big) \nonumber \\
&=&
\begin{cases}x_{\Re}\, h \,E_z  + \tilde{w}_{I}, & \:\:\: c =\hat{c}_{\Re} \\
\tilde{w}_{I}, &   \:\:\: c \not=\hat{c}_{\Re} \end{cases} \\   \label{eq7}
\tilde{r}^k_{Q,c}  &  = & \sum_{l=1}^{L} z_{c,l}\,r_{Q,l}= \sum_{l=1}^{L} z_{c,l}\,\Big(x_{\Im}\,c_{c_{\Im},l}\, h  + w_{Q,l}\Big) \nonumber \\
&=&
\begin{cases}x_{\Im}\, h \,E_z  + \tilde{w}_{Q}, & \:\:\: c =\hat{c}_{\Im} \\
\tilde{w}_{Q}, &   \:\:\: c \not=\hat{c}_{\Im} \end{cases} 
\end{eqnarray}
where, $k$ is an index ranging from 1 to $n_r$, and $c$ is an index ranging from 1 to $2^{N_W}$. The average energy transmitted by each symbol is denoted as $E_z$, which can be expressed as $E_z=\frac{1}{P}\sum_{l=1}^{L}{c^2_{\jmath,l} }$. Additionally, $\tilde{w}_I$ and $\tilde{w}_Q$ are calculated as the $\frac{1}{L}\sum_{l=1}^{L} z_{c,l} w_{I,l}$ and $\frac{1}{L}\sum_{l=1}^{L} z_{c,l} w_{Q,l}$, respectively. The square norm of the received signal vector is computed in order to estimate the spreading code indices for the $I$ and $Q$ components shown in Fig. \ref{system_model}. The spreading code indices that maximize this squared norm value are then determined. It should be noted that orthogonality is exhibited by Walsh codes as $     \sum_{l=1}^{L}z_{a,l}\, z_{b,l}   = \bigg\{\begin{array}{cc}
1, & \,\,\,\,a = b\\ 
0, & \,\,\,\,a \neq b 
\end{array}.$
%follows:
%\begin{equation}
 %    \sum_{l=1}^{L}z_{a,l}\, z_{b,l}   = \bigg\{\begin{array}{cc}
%1, & \,\,\,\,a = b\\ 
%0, & \,\,\,\,a \neq b 
%\end{array}.  
%\end{equation}
In conclusion, the spreading code indices are estimated by analyzing the squared norm of the received signal vector. The estimation involves maximizing this squared norm to determine the spreading code indices that yield the highest value. This is expressed mathematically as follows:
\begin{eqnarray}\label{eq10}
\hat{c}_\Re  &  = & \underset{c}{\mathrm{arg\,max}} \bigg\{\Big|\Big|\tilde{\textbf{r}}_{I,c}\Big|\Big|^2 \bigg\}, \\ \label{eq11}
\hat{c}_\Im  &  = & \underset{c}{\mathrm{arg\,max}} \bigg\{\Big|\Big|\tilde{\textbf{r}}_{Q,c}\Big|\Big|^2 \bigg\}.
\end{eqnarray}
The estimation of spreading code indices prior to the estimation of other parameters offers several benefits, reducing the complexity of the system. By estimating the spreading code indices first, the subsequent estimation of other parameters becomes more efficient. This approach is advantageous because it allows for the use of the selection block at the receiver side to feed back the estimated spreading code indices $(\hat{c}_{\Re}, \hat{c}_{\Im})$ to the set of despreaded vectors. Consequently, only the despreaded data vectors $\tilde{\textbf{r}}_{I,\hat{c}_{\Re}}$ and $\tilde{\textbf{r}}_{Q,\hat{c}_{\Im}}$, associated with the estimated indices, are applied to the input of the ML detector. As a result, unnecessary computations and processing are avoided, significantly reducing the overall complexity of the system. Therefore, by estimating the spreading code indices in advance, the complexity involved in the estimation of other parameters is effectively reduced, leading to improved system efficiency and performance.

After obtaining the estimates $\hat{c}_{\Re}$ and $\hat{c}_{\Im}$ from the maximum index detectors, all possible combinations of the symbol $x$ over the despreaded signals $\tilde{\textbf{r}}_{I,\hat{c}_{\Re}}$ and $\tilde{\textbf{r}}_{Q,\hat{c}_{\Im}}$, associated with the ($\hat{c}_{\Re}$, $\hat{c}_{\Im}$) indices, are considered by the ML detector. Using a maximum likelihood criterion, the detector evaluates each combination to determine the estimations of the symbol $x$. By examining all possible combinations, the detector selects the symbol $x$ values that maximize the likelihood. As a result, the ML detector that estimates the symbol $x$ of the CIM-IT system is defined as follows:
\begin{eqnarray}\label{ML}
\hat{x} & =& \text{arg}\underset{x}{\mathrm{min}} \ \bigg|\bigg|\Big(\tilde{\textbf{r}}_{I,\hat{c}_\Re}+j\tilde{\textbf{r}}_{Q,\hat{c}_\Im}\Big)-E_z x \mathbf{h} \bigg|\bigg|^2.  
\end{eqnarray}
By employing the bit demapper technique, the receiver utilizes the detected values $(\hat{x}, \hat{c}_{\Re}, \hat{c}_{\Im})$ to reconstruct the originally transmitted bit sequence $\hat{\textbf{b}}$. This reconstruction process occurs at the receiver of the CIM-IT system, aided by a demapper in Fig. \ref{system_model}. The subsequently obtained estimated bit sequence, denoted as $\hat{\textbf{b}}$, is converted back into pixel values. By utilizing these pixels, a series of enhancement filters are applied to the image, resulting in the reconstruction of the transmitted image at the receiver.

\begin{figure}[t]
\centering{\includegraphics[scale=0.5]{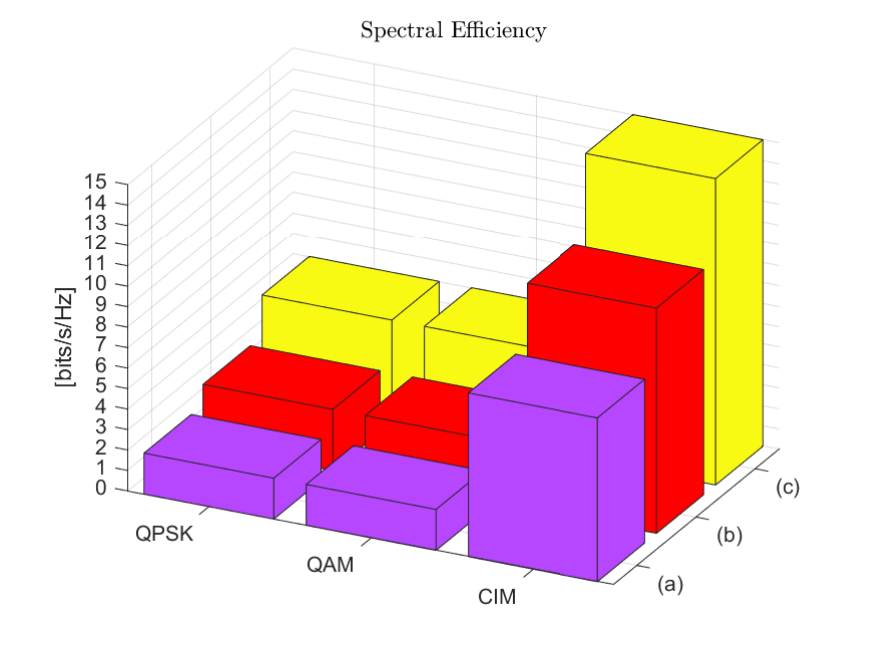}}
	\caption{The spectral efficiency comparisons for the CIM, QAM, and QPSK systems with the following system parameters: (a) $M=4$, $N_W=3$, (b) $M=8$, $N_W=4$, and (c) $M=32$, $N_W=5$ }
	\label{cimspectral} 		
\end{figure}

\section{Performance Analysis}
In this section, comparative analyses are conducted to assess the performance of the CIM scheme in relation to the QAM and QPSK schemes. Important performance indicators, including spectral efficiency, throughput, and energy efficiency, are utilized for evaluation purposes. The objective of these analyses is to determine how the CIM system fares in comparison to the other systems across these critical aspects.

Spectral efficiency measures the amount of information that can be conveyed over a given bandwidth. Fig. \ref{cimspectral} presents a comparative analysis of the spectral efficiency of the CIM, QAM, and QPSK schemes. For CIM, QAM, and QPSK systems, the system parameters ($M=4$, $N_W=3$), ($M=4$), and ($M=4$) for (a), ($M=8$, $N_W=4$), ($M=8$), and ($M=8$) for (b) and finally ($M=32$, $N_W=5$), ($M=32$) and ($M=32$) for (c) are used respectively. The comparisons presented in Fig. \ref{cimspectral} clearly demonstrate that the CIM system outperforms other systems in terms of spectral efficiency. The CIM system's higher spectral efficiency enables faster transmission speeds, reducing the time required to send or receive images. This is particularly advantageous for real-time applications such as video streaming or video conferencing, where low latency and seamless playback are crucial. Also, high spectral efficiency enables more data transmission over limited bandwidth. It allows for faster transmission of higher-resolution images.

\begin{figure}[t]
\centering{\includegraphics[scale=0.5]{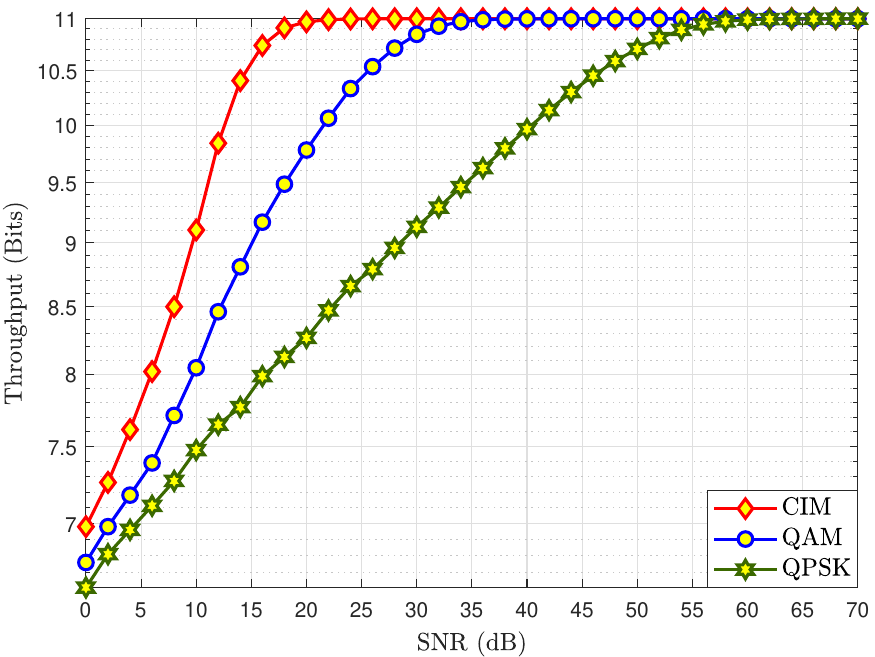}}
	\caption{The throughput comparisons of CIM, QAM, and QPSK systems for $\eta=11$ and $n_r=3$.}
	\label{throughput1} 		
\end{figure}

\begin{figure}[t]
\centering{\includegraphics[scale=0.46]{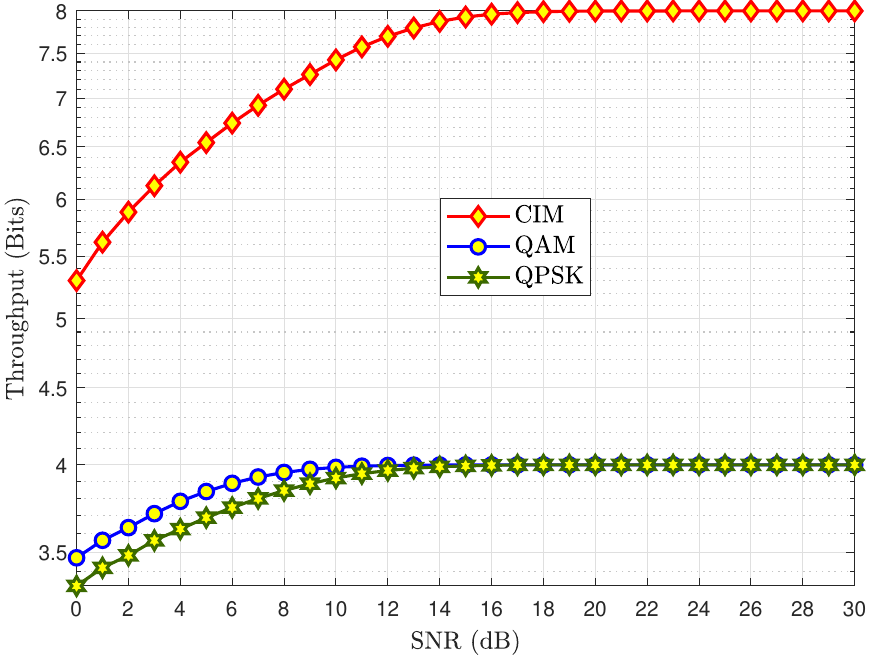}}
	\caption{The throughput comparisons of CIM, QAM, and QPSK systems for $M=16$, $N_W=2$, $n_r=5$.}
	\label{throughput2} 		
\end{figure}

The rate at which data can be reliably transmitted and received over a wireless channel is measured by throughput, a crucial metric in wireless communication networks. It pertains to the receiver's ability to accurately receive data bits from the transmitted ones. Throughput quantifies the successful transmission and reception of data bits within a specified time frame, considering any potential errors or losses during transmission. Several factors, such as channel conditions, the utilized modulation scheme, and the SNR of the transmitted signal, impact throughput. The throughput of the CIM system can be computed as follows \cite{Tse}:
\begin{equation}\label{throughput}
\mathcal{T}_{\text{CIM}}    =  \frac{\big(1-\text{ABER}_{\text{CIM}}\big)}{T_s}\eta,
\end{equation}  
where the probability of the receiver accurately capturing the transmitted bits within the symbol transmission duration $T_s$ in the CIM system is represented as $(1-\text{ABER}_{\text{CIM}})$. This indicates the probability of correctly obtaining the transmitted bits. Throughput comparisons of CIM, QAM, and QPSK systems are presented in Fig. \ref{throughput1} for $\eta=11$ and $n_r=3$. CIM, QAM, and QPSK systems use system parameters ($M=8$, $N_W=4$, $L=32$), ($M=2048$), and ($M=2048$) respectively. Fig. \ref{throughput1} clearly shows that the CIM system achieves higher throughput at lower SNR than QAM and QPSK systems. A benchmark including a similar throughput comparison of CIM, QAM, and QPSK systems is also presented in Fig. \ref{throughput2} for $n_r=5$. Here, system parameters $M=16$, $N_W=2$, and $L=16$ are selected. For the relevant system parameters, it is easily seen in Fig. \ref{throughput2} that the CIM system provides higher efficiency than the other systems. When the CIM system achieves high throughput, several benefits are obtained in terms of image transfer. By associating each bit with a pixel, the CIM system can efficiently transmit a large amount of image data within a given time frame. The high throughput ensures that a significant number of data bits, representing pixels, can be accurately transmitted and received. This results in faster and more reliable image transfer, allowing for real-time or near-real-time transmission of high-quality images. Furthermore, the high throughput capability of the CIM system enables the transmission of complex and detailed images without compromising their quality. 

\begin{figure}[t]
\hspace*{-1.4cm} 
\centering{\includegraphics[scale=0.35]{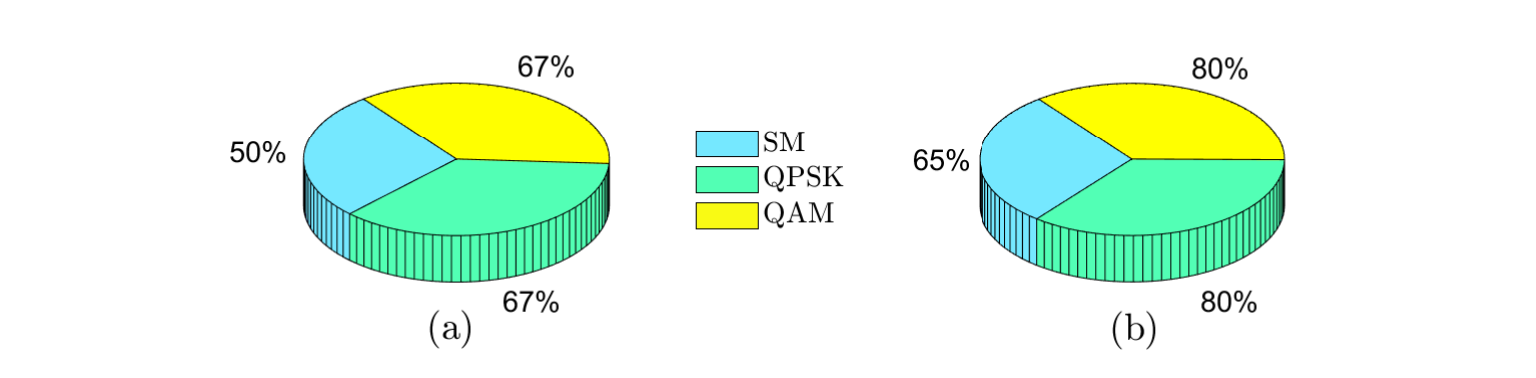}}
	\caption{For the system parameters (a) ($M=4$, $N_W=2$, $N_T=2$)  and (b) ($M=16$, $N_W=8$, $N_T=8$), the energy savings percentages of the proposed CIM system compared to SM, QAM, and QPSK systems. }
	\label{energycim} 		
\end{figure}

Energy efficiency is considered to be of great importance in today's world due to several reasons. Firstly, the increasing demand for energy and the limited availability of energy resources necessitate the need for improved energy efficiency. More efficient systems enable the more effective utilization of existing energy sources, thereby reducing energy consumption and conserving resources. Additionally, energy efficiency helps mitigate environmental impacts. Energy production processes, such as the combustion of fossil fuels, contribute to greenhouse gas emissions and climate change. By focusing on energy efficiency, less energy is consumed, resulting in reduced greenhouse gas emissions and enhanced environmental sustainability. Therefore, energy-efficient technologies like the CIM system play a significant role today by minimizing energy consumption, preserving resources, mitigating environmental impacts, and providing economic advantages. In the CIM system, the majority of the information bits are mapped in the active spreading code indices. By utilizing indices to transmit information bits, the system minimizes or completely eliminates the energy consumption associated with data transmission. As a result, the CIM system achieves remarkable energy efficiency. The energy-saving percentage ($\text{E}_{sav}$) of the CIM system, compared to other benchmark systems per $\eta$ bits, can be expressed as follows:
\begin{equation}\label{energy_cim}
E_{\text{sav}}=\Big(1-\frac{n_c}{\eta}\Big)E_b\%,
\end{equation}
where $n_c$ represents the spectral efficiency of the compared systems while $E_b$ represents the bit energy. Fig. \ref{energycim} presents the percentage of energy efficiency provided by the CIM system in comparison to the benchmark systems including SM, QAM, and QPSK. In the comparison presented in Fig. \ref{energycim} (a), the system parameters ($M=4$, $N_W=2$, $N_T=2$) are used, while in Fig. \ref{energycim} (b), the system parameters ($M=16$, $N_W=8$, $N_T=8$) are used. Higher energy efficiency is observed with the CIM system compared to all the benchmark systems. In Fig. \ref{energycim} (a), the CIM system yields energy efficiency that is $50\%$, $67\%$, and $67\%$ higher than the SM, QAM, and QPSK systems, respectively. Similarly In Fig. \ref{energycim} (b), the CIM system provides energy efficiency that is $65\%$, $80\%$, and $80\%$ higher than the SM, QAM, and QPSK systems, respectively. The high energy efficiency of the CIM system, achieved through low energy consumption during bit transmission, provides significant benefits for image transfer, including extended battery life, enhanced mobility, and reduced heat generation.

\section{Image denoising based enhancement methods}
In the proposed CIM-IT system, the original black-and-white image is transmitted over the Rayleigh wireless channel. It is then subjected to the effects of the wireless channel such as reflection, scattering, diffraction, etc., and then further degraded by the addition of Gaussian noise.  As a result, it is necessary to improve the noisy image at the receiver of the CIM-IT system. The noise in the image is similar to salt and pepper noise. In this section, the enhancement filters applied to the noisy image obtained at the receiver of the proposed CIM-IT system such as the non-local means (NLM) filter, Wiener filter, wavelet filter, morphological filter, majority filter, and median filter \cite{Gonzalez2002}. Each enhancement filter's theoretical background is discussed, and its effectiveness in removing noise from noisy images is analyzed.

\subsection{Non-Local Means Filter}

NLM filtering is a denoising technique that estimates the value of a pixel by taking the weighted average of similar pixels found within a larger search window \cite{Buades2005}. Unlike traditional filters that rely on local neighborhoods, NLM utilizes redundancy in natural images by averaging pixels with similar structures, even if they are far apart. This approach is particularly effective in preserving textures and fine details while reducing noise.

%NLM has been proven to be highly effective in denoising images with low to moderate noise levels while preserving important image structures \cite{Buades2005}. However, its computational complexity can be high, making it less suitable for real-time applications without optimizations.

\subsection{Wiener Filter}
The Wiener filter is a linear filtering technique that minimizes the mean square error between the original and estimated images. It is particularly effective in scenarios where the noise has a known statistical property and is expressed mathematically as follows \cite{Lee1980}:
\begin{equation}
G(u,v) = \frac{H^*(u,v) S_X(u,v)}{|H(u,v)|^2 S_X(u,v) + S_N(u,v)},
\end{equation}
where \( G(u,v) \) is the output image in the frequency domain, \( H(u,v) \) is the degradation function, \( S_X(u,v) \) and \( S_N(u,v) \) are the power spectra of the original image and noise, respectively.

\subsection{Wavelet Filter}
Wavelet filter decomposes the noisy image into different frequency components and processes them individually. This technique allows noise removal while preserving significant image features \cite{Mallat1989}. The denoising process begins by decomposing the image using a wavelet transform, which breaks down the image into various frequency components at different scales. Next, a threshold is applied to the wavelet coefficients, selectively reducing those with small magnitudes, as they are likely to represent noise. Finally, the image is reconstructed using the inverse wavelet transform, leveraging the sparsity of the wavelet coefficients to preserve essential features while minimizing noise.

\subsection{Morphological Operations}
Morphological operations are fundamental tools in image processing, especially for binary and grayscale images. These operations are based on set theory and are used to extract meaningful structures, suppress noise, and refine image features. The primary morphological operations include erosion, dilation, opening, and closing \cite{Gonzalez2002}.

\textbf{1. Erosion:} 
Erosion is an operation that shrinks the objects in a binary image by removing pixels on object boundaries. It is mathematically defined as:
\begin{equation}
I \ominus B = \{z \mid B_z \subseteq I\},
\end{equation}
where \( I \) is the input image, \( B \) is the structuring element, and \( B_z \) represents the translation of \( B \) by \( z \). Also, here, \( \ominus \) denotes the erosion operation. The result of erosion is a new image where the boundaries of objects have been eroded.

\textbf{2. Dilation:}
Dilation is the dual operation of erosion, which adds pixels to the boundaries of objects. It is mathematically defined as:
\begin{equation}
I \oplus B = \{z \mid B_z \cap I \neq \emptyset\},
\end{equation}
here, \( \oplus \) denotes the dilation operation. Dilation enlarges the objects in the image, filling in small holes and gaps.

\textbf{3. Opening:}
The opening is a sequence of erosion followed by a dilation. This operation is useful for removing small objects and noise. It is defined as:
\begin{equation}
I \circ B = (I \ominus B) \oplus B,
\end{equation}
where \( \circ \) denotes the opening operation. Opening smooths the contour of an object, breaking narrow isthmuses and eliminating thin protrusions.

\textbf{4. Closing:}
Closing is the dual operation of opening, consisting of a dilation followed by an erosion. It is defined as:
\begin{equation}
I \bullet B = (I \oplus B) \ominus B,
\end{equation}
where \( \bullet \) denotes the closing operation. Closing is useful for closing small holes and connecting nearby objects in an image.

Morphological operations are particularly effective for removing salt-and-pepper noise in binary images. The opening operation removes small isolated noise points, while the closing operation fills small gaps and holes. When applied sequentially, these operations can significantly reduce noise and improve image quality.

\subsection{Majority Filter}
Majority filtering is a simple yet effective technique for binary images. It replaces each pixel with the most common value among its neighbors. This method is particularly useful for removing isolated noise pixels in binary images \cite{Gonzalez2002, Serra1982}. Mathematically, it can be expressed as:
\begin{equation}
I'(i,j) = \text{mode}\{I(i+k,j+l) \mid -m \leq k,l \leq m \},
\end{equation}
where the mode function determines the most frequent value within the neighborhood. Also, In the expression, $i$ represents the row index, and $j$ represents the column index of a pixel in the image, with ($i$, $j$) indicating the position of the pixel within the 2D grid.

\subsection{Median Filter}
Median filtering is a non-linear technique often used to remove salt-and-pepper noise. The median filter replaces each pixel value with the median value of its neighboring pixels. This method effectively preserves edges while removing noise, as described by Gonzalez and Woods \cite{Gonzalez2002}. The median filter can be described mathematically as follows:
\begin{equation}
I'(i,j) = \text{median}\{I(i+k,j+l) \mid -m \leq k,l \leq m \},
\end{equation}
where \( I'(i,j) \) is the denoised pixel value, and the operation is applied to a \( (2m+1) \times (2m+1) \) neighborhood.

\section{Simulation Results}

The simulation results section presents the findings of the conducted simulations on image transmission over a wireless Rayleigh fading channel using the CIM technique. Each bit is represented by a single pixel, and the simulations focused on investigating the impact of signal-to-noise ratio (SNR) on the bit error rate (BER) performance. 
%A comprehensive set of graphs is generated to visualize the BER performance under different SNR conditions. 
Additionally, the BER performance of the CIM system is compared to that of traditional wireless communication systems, namely quadrature amplitude modulation (QAM) and quadrature phase shift keying (QPSK). 
%The results of the comparisons are thoroughly discussed and analyzed. The results revealed the advantages and limitations of the CIM approach in terms of BER performance, facilitating a comprehensive discussion of its effectiveness compared to established techniques. Furthermore, the impact of CIM system parameters on the BER performance is examined by altering these parameters during the simulations. The effects of parameters such as code length, spreading codes, and modulation order are investigated. Multiple graphs are generated to illustrate the variations in BER performance resulting from parameter adjustments, providing insights into the sensitivity of the system to different parameter configurations. 
Also, the expression for SNR in simulations is defined as $\mathrm{SNR (dB)}$ being equal to $10\log_{10}(E_s/N_0)$. In all simulation results, it is assumed that the fading channels are characterized by uncorrelated Rayleigh distribution, and the Monte Carlo method is utilized.

\begin{figure}[t]
\centering{\includegraphics[scale=0.5]{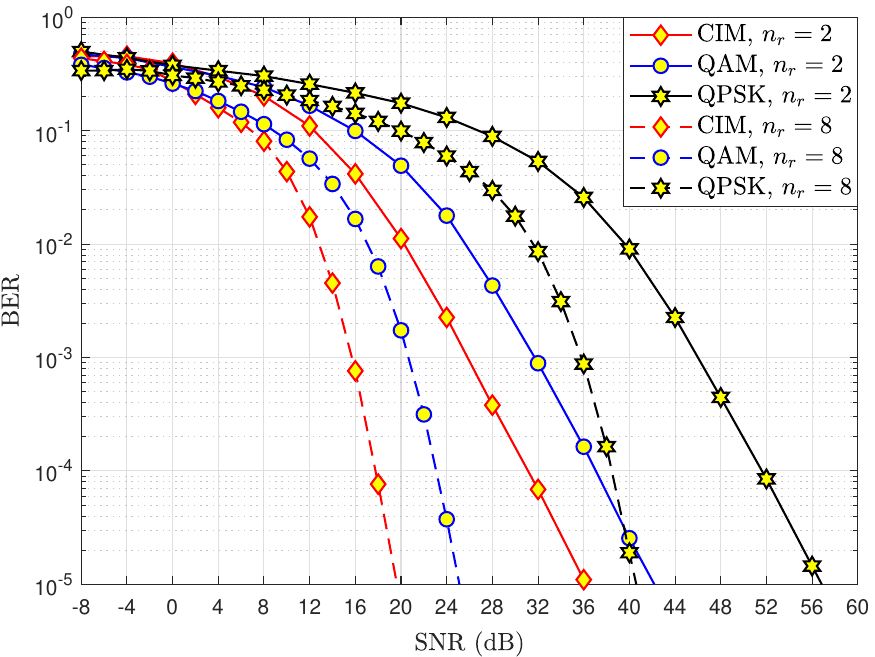}}
	\caption{Performance comparisons of CIM, QAM, and QPSK systems for $n_r=2$ and $n_r=8$ while $\eta=8$.}
	\label{benckmarknr28} 		
\end{figure}

In Fig. \ref{benckmarknr28}, the BER comparisons are conducted between the CIM system and traditional wireless communication systems such as QAM and QPSK, considering different numbers of receiver antennas. The system parameters used in the CIM, QAM, and QPSK systems are denoted as ($M=16$, $N_W=2$, $L=16$), ($M=256$), and ($M=256$), respectively. The results indicate significant improvements in the BER performance of all systems as the number of receiver antennas increases. Furthermore, it is evident that the CIM system exhibits superior performance compared to QAM and QPSK systems for the same spectral efficiency. In the different system parameters, the results of the BER comparisons between the CIM system, and traditional wireless communication systems like QAM and QPSK indicate that the CIM system excels in image transmission by delivering enhanced image quality, improved reliability, and superior overall performance.

\begin{figure}[t]
\centering{\includegraphics[scale=0.5]{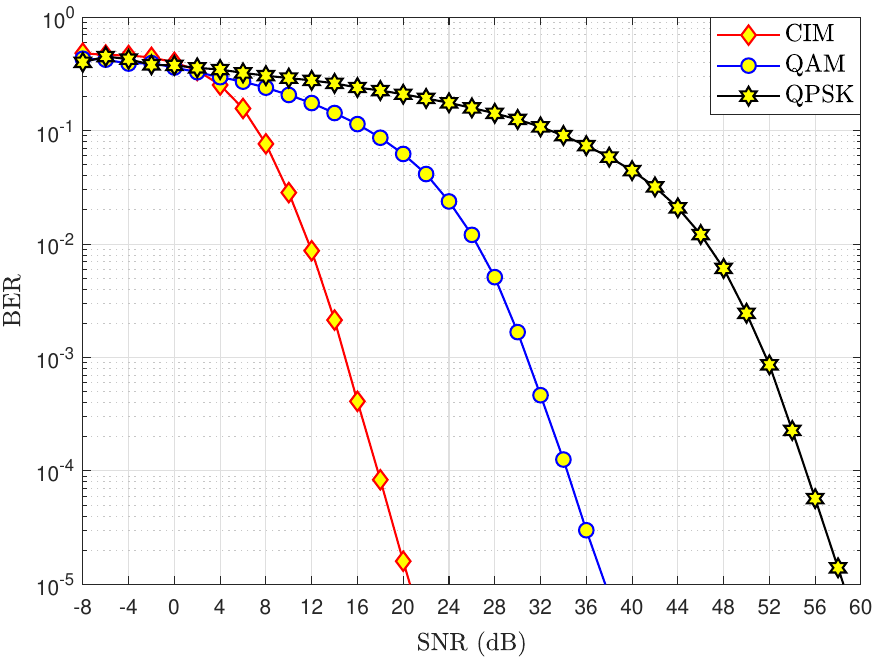}}
	\caption{Performance comparisons of CIM, QAM, and QPSK systems for $n_r=4$ while $\eta=10$.}
	\label{benckmarknr4} 		
\end{figure}

In a similar manner to Fig. \ref{benckmarknr28}, for a different case, a comparative BER analysis is conducted to evaluate the performance of the CIM system in relation to QAM and QPSK systems in Fig. \ref{benckmarknr4}. The CIM, QAM, and QPSK systems use specific system parameters, represented as ($M=4$, $N_W=4$, $L=64$), ($M=1024$), and ($M=1024$), respectively. Where it is crucial to highlight that the CIM system outperforms substantially other systems compared to the scenario depicted in Fig. \ref{benckmarknr28}. This is primarily due to the efficient transmission of the majority of data bits through spreading code indices.  This finding underscores the CIM system's remarkable performance in image transmission, as it yields improved BER and spectral efficiency results, while effectively utilizing spreading code indices to carry the majority of the data bits.

\begin{figure}[t]
\centering{\includegraphics[scale=0.5]{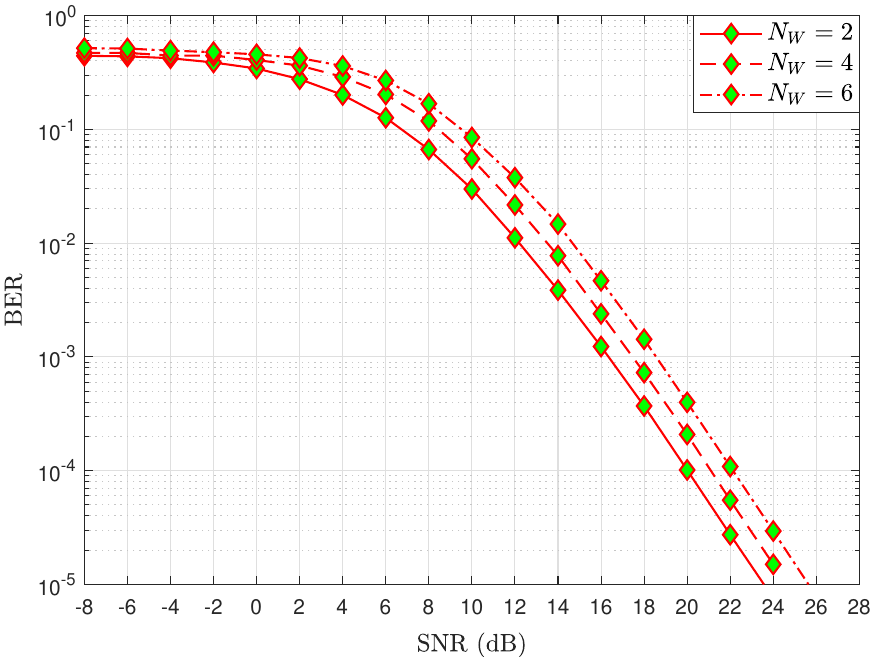}}
	\caption{Performance comparisons of CIM system for  $M=4$, $L=128$ and $n_r=3$}
	\label{cimncsms} 		
\end{figure}

The BER performance of the CIM system for parameters $N_W=2,4,6$, $M=4$, $L=128$, and $n_r=3$ is presented in Fig. \ref{cimncsms}. Fig. \ref{cimncsms} focuses on examining how the BER performance of the CIM system is influenced by variations in the $N_W$ and $M$ parameters. The results reveal that increasing the $N_W$ and $M$ parameters improves spectral efficiency, facilitating faster data transmission. However, this enhancement comes at the cost of degradation in the BER performance, leading to an increased number of erroneous transmitted bits. These findings emphasize the significance of this analysis in the context of wireless image transfer. They underscore the importance of carefully selecting the $N_W$ and $M$ parameters to strike a balance between spectral efficiency and BER performance, ensuring optimal image quality and reliable communication in wireless systems.

\begin{figure}[t]
\centering{\includegraphics[scale=0.5]{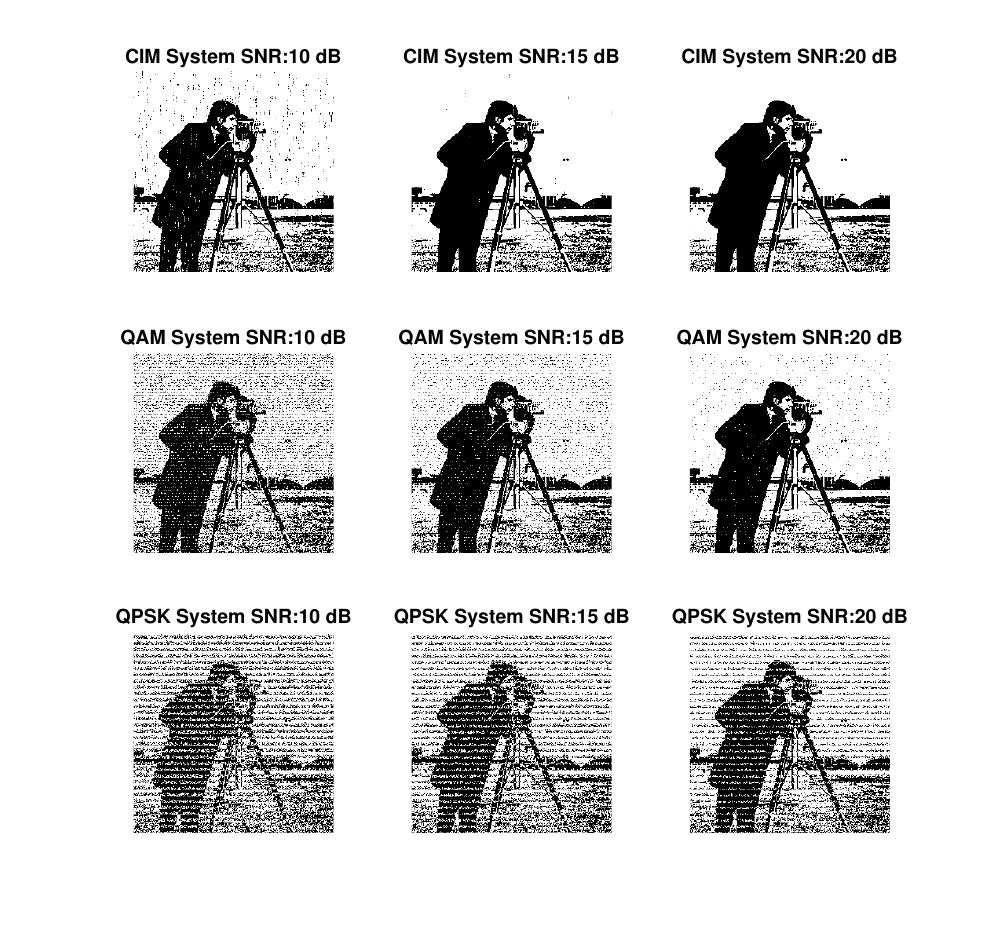}}
	\caption{Images obtained at the receiver by transmitting with CIM, QAM, and QPSK systems according to changing SNR values while $n_r=4$.}
	\label{imageSNRs} 		
\end{figure}

\begin{figure}[t]
\centering{\includegraphics[scale=0.485]{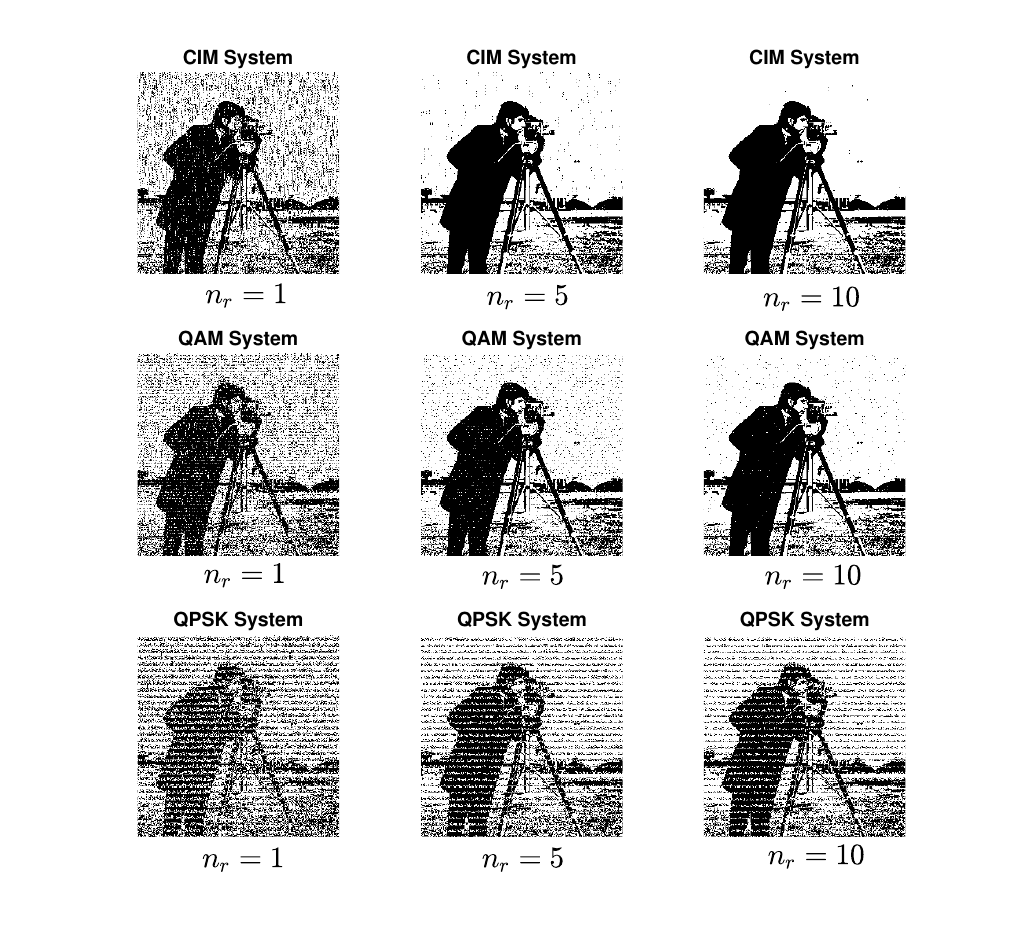}}
	\caption{Images obtained at the receiver by transmitting with CIM, QAM, and QPSK systems according to changing $n_r$ values while $SNR=16$ dB.}
	\label{imageNRs} 		
\end{figure}

Fig. \ref{imageSNRs} shows the degraded images obtained at the receiver when wireless image transmission is performed with CIM, QAM, and QPSK systems for $SNR=(10,15,20) \ \text{dB}$ while $n_r=4$ and $\eta=8$. CIM, QAM, and QPSK systems use system parameters ($M=4$, $N_W=3$, $L=32$), ($M=256$), and ($M=256$), respectively. The results obtained indicated that the CIM system provides a reduced distortion and lower BER during image transmission. It is observed that as the SNR value increased, the quality of the transmitted image improved, resulting in a decrease in the number of erroneously transmitted bits. Also, the BER values for CIM, QAM, and QPSK systems are ($2.1\times 10^{-2}$, $6.7\times 10^{-4}$, $0$), ($1.2\times 10^{-1}$, $5.7\times 10^{-2}$, $1.3\times 10^{-2}$) and ($2.4\times 10^{-1}$, $1.9\times 10^{-1}$, $1.3\times 10^{-1}$) for $SNR=(10,15,20)$ dB, respectively. As a result, it is observed that as the SNR value increases, the transmitted images are subject to less distortion and it is shown that the CIM system can transmit much higher quality images than QAM and QPSK systems for the same conditions.

The degraded images obtained at the receiver during wireless image transmission with CIM, QAM, and QPSK systems are shown in Fig. \ref{imageNRs} for $SNR=16$ dB. The experiments are conducted with $n_r=(1, 5, 10)$ while considering $\eta=8$. The system parameters for CIM, QAM, and QPSK systems were set as follows: ($M=16$, $N_W=2$, $L=64$) for CIM, $M=256$ for QAM, and $M=256$ for QPSK. The results indicate that the CIM system exhibits reduced distortion and lower BER during image transmission compared to QAM and QPSK systems. With increasing $n_r$ values, the quality of the transmitted image improves, leading to a decrease in the number of erroneously transmitted bits. The BER values for CIM, QAM, and QPSK systems at $n_r=(1, 5, 10)$ are found to be ($1.3\times 10^{-1}$, $6.3\times 10^{-3}$, $4.0\times 10^{-4}$), ($1.6\times 10^{-1}$, $3.5\times 10^{-2}$, $8.8\times 10^{-3}$), and ($2.6\times 10^{-1}$, $1.6\times 10^{-1}$, $1.3\times 10^{-1}$), respectively. These results highlight that as the $n_r$ value increases, the distortion in transmitted images decreases. Moreover, the CIM system outperforms the QAM and QPSK systems under the same conditions, enabling the transmission of significantly higher-quality images.

\begin{figure}[t!]
    \centering
    \begin{subfigure}[b]{0.11\textwidth}
        \centering
        \includegraphics[width=\textwidth]{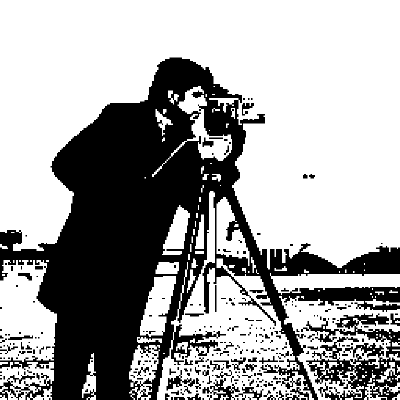}
        %\caption{Original image \newline}
        \caption{}
        \label{org_image}
    \end{subfigure}
    \hfill
    \begin{subfigure}[b]{0.11\textwidth}
        \centering
        \includegraphics[width=\textwidth]{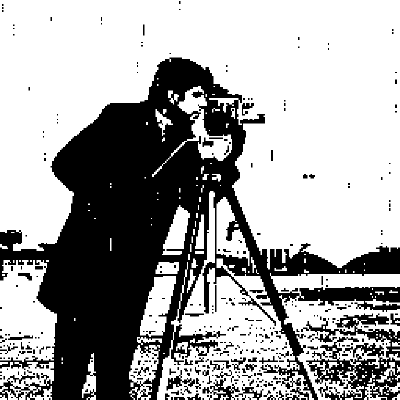}
        %\caption{Noisy image at the receiver of the CIM-IT}
        \caption{}
        \label{case1_noisy}
    \end{subfigure}
    \hfill
    \begin{subfigure}[b]{0.11\textwidth}
        \centering
        \includegraphics[width=\textwidth]{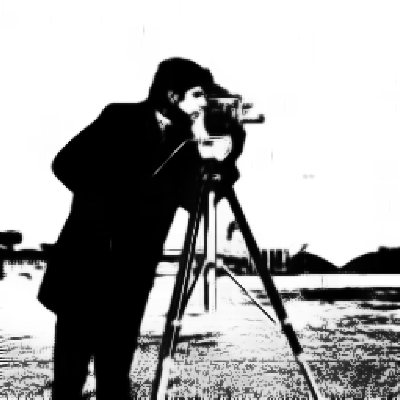}
        %\caption{Image enhanced with NLM denoising technique}
        \caption{}
        \label{case1_enh1}
    \end{subfigure}
    \hfill
    \begin{subfigure}[b]{0.11\textwidth}
        \centering
        \includegraphics[width=\textwidth]{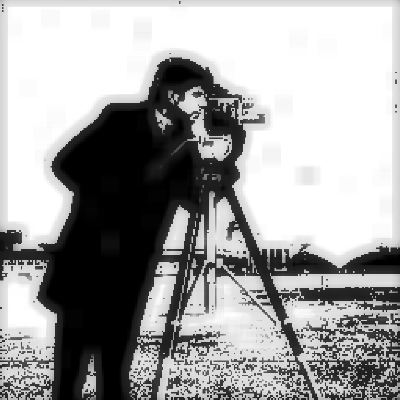}
        %\caption{Image enhanced with Wiener filter}
        \caption{}
        \label{case1_enh2}
    \end{subfigure}
    \hfill
    \begin{subfigure}[b]{0.11\textwidth}
        \centering
        \includegraphics[width=\textwidth]{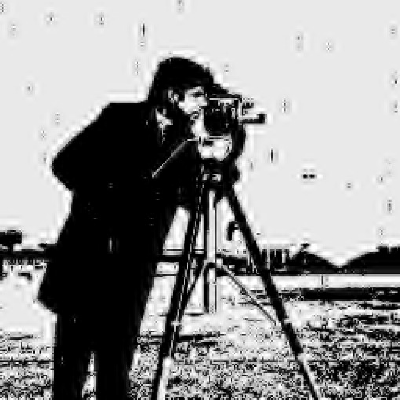}
       % \caption{Image enhanced with wavelet filter}
       \caption{}
        \label{case1_enh3}
    \end{subfigure}
    \hfill
    \begin{subfigure}[b]{0.11\textwidth}
        \centering
        \includegraphics[width=\textwidth]{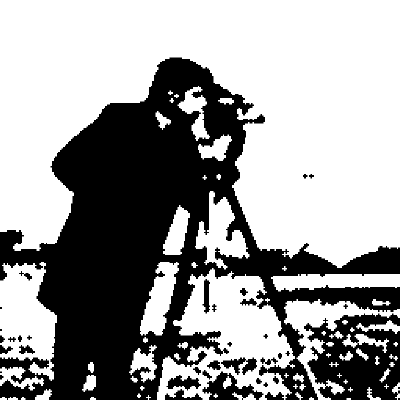}
        %\caption{Image enhanced with morphological filter}
        \caption{}
        \label{case1_enh4}
    \end{subfigure}
    \hfill
    \begin{subfigure}[b]{0.11\textwidth}
        \centering
        \includegraphics[width=\textwidth]{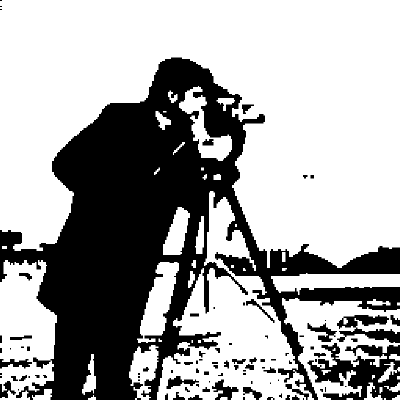}
        %\caption{Image enhanced with majority filter}
        \caption{}
        \label{case1_enh5}
    \end{subfigure}
    \hfill
    \begin{subfigure}[b]{0.11\textwidth}
        \centering
        \includegraphics[width=\textwidth]{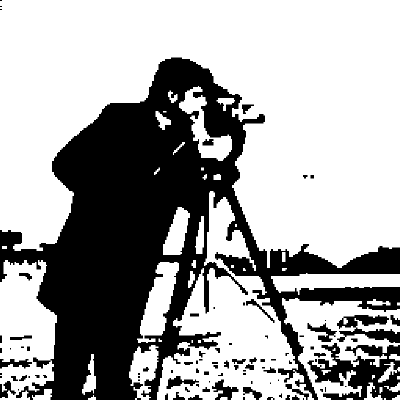}
        %\caption{Image enhanced with median filter}
        \caption{}
        \label{case1_enh6}
    \end{subfigure}
    \caption{Image of the noise at the receiver of the CIM system and images enhanced with different techniques for system parameters  $M=4$, $N_W=3$, $L=16$, $n_r=2$, and $SNR=20$ dB. (a) original image, (b) noisy image at the receiver of the CIM-IT, (c) image enhanced with NLM denoising technique, (d) image enhanced with Wiener filter, (e) image enhanced with wavelet filter, (f) image enhanced with morphological filter, (g) image enhanced with majority filter, and (h) image enhanced with median filter.}
    \label{case1_all}
\end{figure}

\begin{figure}[t!]
    \centering
    \begin{subfigure}[b]{0.11\textwidth}
        \centering
        \includegraphics[width=\textwidth]{enh_files/org_image.pdf}
        %\caption{Original image \newline}
        \caption{}
        \label{org_image2}
    \end{subfigure}
    \hfill
    \begin{subfigure}[b]{0.11\textwidth}
        \centering
        \includegraphics[width=\textwidth]{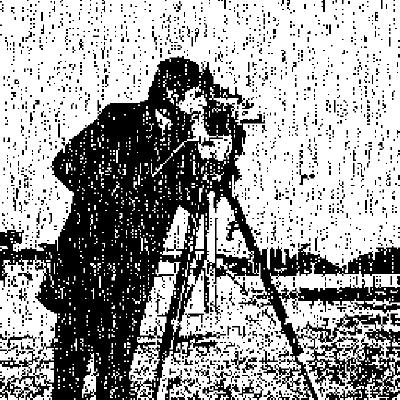}
        %\caption{Noisy image at the receiver of the CIM-IT}
        \caption{}
        \label{case2_noisy}
    \end{subfigure}
    \hfill
    \begin{subfigure}[b]{0.11\textwidth}
        \centering
        \includegraphics[width=\textwidth]{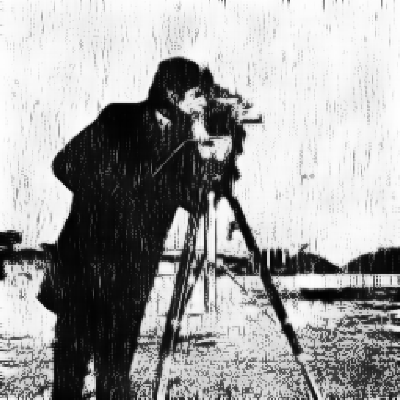}
        %\caption{Image enhanced with NLM denoising technique}
        \caption{}
        \label{case2_enh1}
    \end{subfigure}
    \hfill
    \begin{subfigure}[b]{0.11\textwidth}
        \centering
        \includegraphics[width=\textwidth]{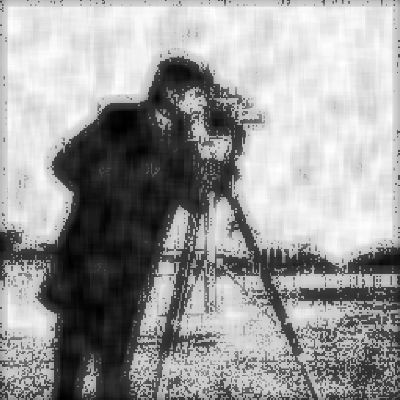}
       % \caption{Image enhanced with Wiener filter}
        \caption{}
        \label{case2_enh2}
    \end{subfigure}
    \hfill
    \begin{subfigure}[b]{0.11\textwidth}
        \centering
        \includegraphics[width=\textwidth]{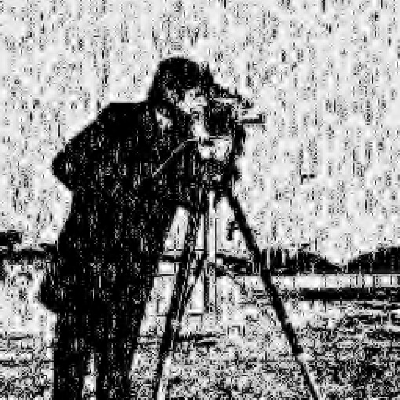}
        %\caption{Image enhanced with wavelet filter}
        \caption{}
        \label{case2_enh3}
    \end{subfigure}
    \hfill
    \begin{subfigure}[b]{0.11\textwidth}
        \centering
        \includegraphics[width=\textwidth]{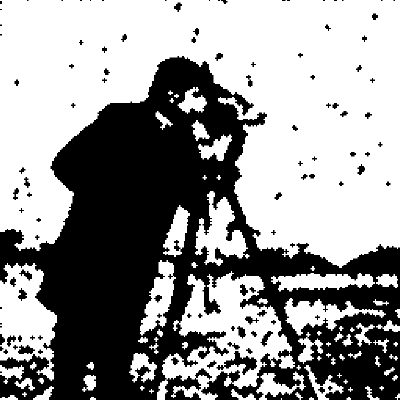}
        %\caption{Image enhanced with morphological filter}
        \caption{}
        \label{case2_enh4}
    \end{subfigure}
    \hfill
    \begin{subfigure}[b]{0.11\textwidth}
        \centering
        \includegraphics[width=\textwidth]{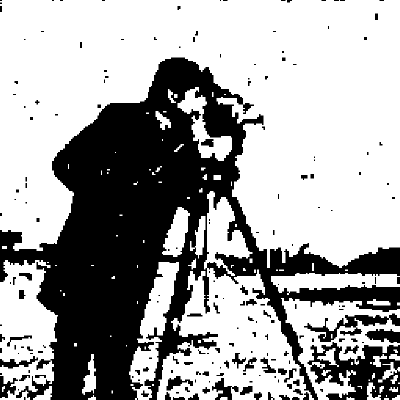}
        %\caption{Image enhanced with majority filter}
        \caption{}
        \label{case2_enh5}
    \end{subfigure}
    \hfill
    \begin{subfigure}[b]{0.11\textwidth}
        \centering
        \includegraphics[width=\textwidth]{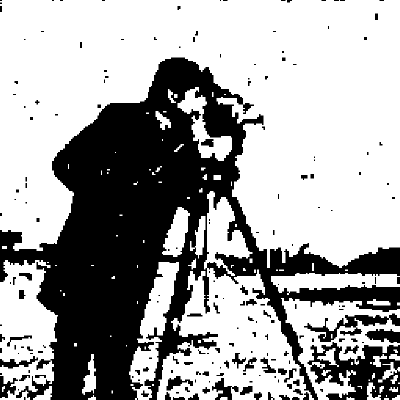}
        %\caption{Image enhanced with median filter}
        \caption{}
        \label{case2_enh6}
    \end{subfigure}
    \caption{Image of the noise at the receiver of the CIM system and images enhanced with different techniques for system parameters  $M=4$, $N_W=3$, $L=16$, $n_r=2$, and $SNR=10$ dB. (a) original image, (b) noisy image at the receiver of the CIM-IT, (c) image enhanced with NLM denoising technique, (d) image enhanced with Wiener filter, (e) image enhanced with wavelet filter, (f) image enhanced with morphological filter, (g) image enhanced with majority filter, and (h) image enhanced with median filter.}
    \label{case2_all}
\end{figure}
\begin{figure}[t!]
    \centering
    \begin{subfigure}[b]{0.11\textwidth}
        \centering
        \includegraphics[width=\textwidth]{enh_files/org_image.pdf}
        %\caption{Original image \newline}
        \caption{}
        \label{org_image3}
    \end{subfigure}
    \hfill
    \begin{subfigure}[b]{0.11\textwidth}
        \centering
        \includegraphics[width=\textwidth]{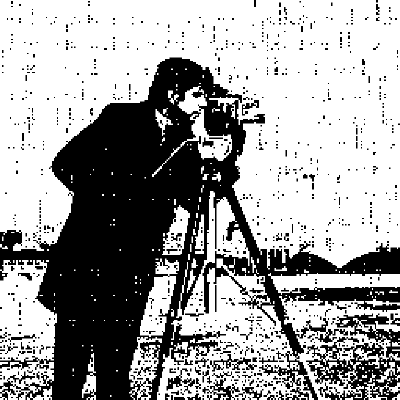}
        %\caption{Noisy image at the receiver of the CIM-IT}
        \caption{}
        \label{case3_noisy}
    \end{subfigure}
    \hfill
    \begin{subfigure}[b]{0.11\textwidth}
        \centering
        \includegraphics[width=\textwidth]{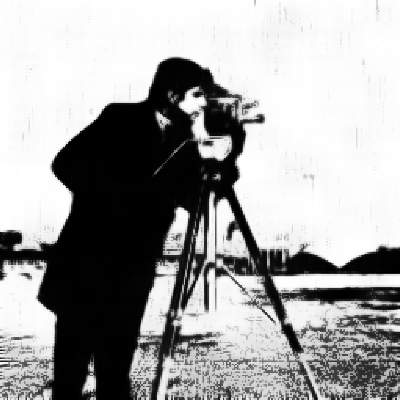}
        %\caption{Image enhanced with NLM denoising technique}
        \caption{}
        \label{case3_enh1}
    \end{subfigure}
    \hfill
    \begin{subfigure}[b]{0.11\textwidth}
        \centering
        \includegraphics[width=\textwidth]{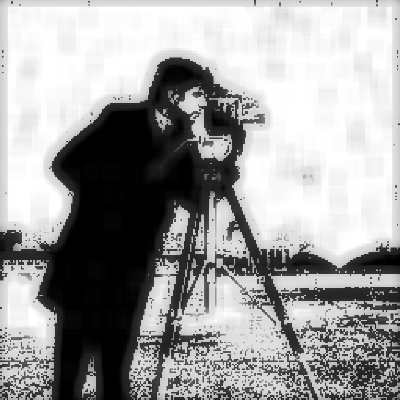}
        %\caption{Image enhanced with Wiener filter}
        \caption{}
        \label{case3_enh2}
    \end{subfigure}
    \hfill
    \begin{subfigure}[b]{0.11\textwidth}
        \centering
        \includegraphics[width=\textwidth]{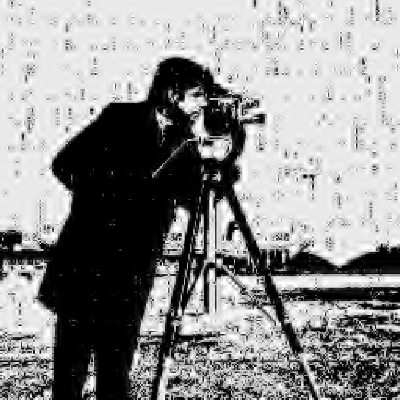}
        %\caption{Image enhanced with wavelet filter}
        \caption{}
        \label{case3_enh3}
    \end{subfigure}
    \hfill
    \begin{subfigure}[b]{0.11\textwidth}
        \centering
        \includegraphics[width=\textwidth]{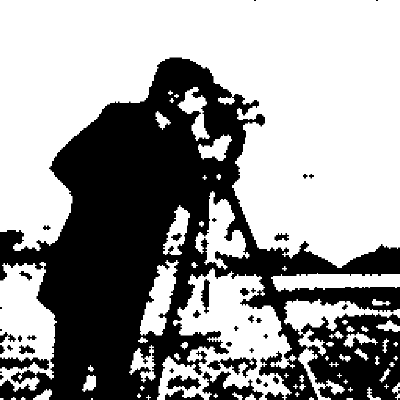}
        %\caption{Image enhanced with morphological filter}
        \caption{}
        \label{case3_enh4}
    \end{subfigure}
    \hfill
    \begin{subfigure}[b]{0.11\textwidth}
        \centering
        \includegraphics[width=\textwidth]{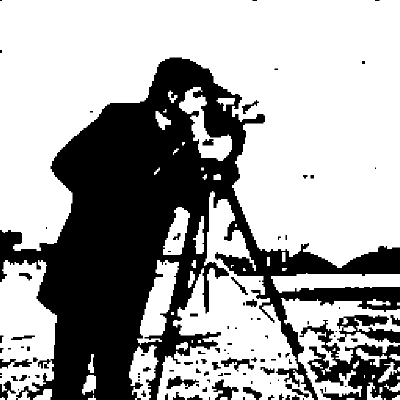}
        %\caption{Image enhanced with majority filter}
        \caption{}
        \label{case3_enh5}
    \end{subfigure}
    \hfill
    \begin{subfigure}[b]{0.11\textwidth}
        \centering
        \includegraphics[width=\textwidth]{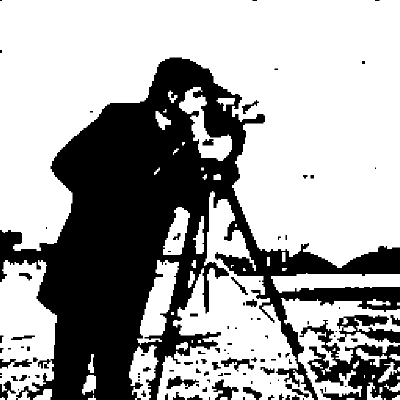}
        %\caption{Image enhanced with median filter}
        \caption{}
        \label{case3_enh6}
    \end{subfigure}
    \caption{Image of the noise at the receiver of the CIM system and images enhanced with different techniques for system parameters  $M=64$, $N_W=5$, $L=128$, $n_r=2$, and $SNR=20$ dB. (a) original image, (b) noisy image at the receiver of the CIM-IT, (c) image enhanced with NLM denoising technique, (d) image enhanced with Wiener filter, (e) image enhanced with wavelet filter, (f) image enhanced with morphological filter, (g) image enhanced with majority filter, and (h) image enhanced with median filter.} 
    \label{case3_all}
\end{figure}

\begin{figure}[t!]
    \centering
    \begin{subfigure}[b]{0.11\textwidth}
        \centering
        \includegraphics[width=\textwidth]{enh_files/org_image.pdf}
        %\caption{Original image \newline}
        \caption{}
        \label{org_image4}
    \end{subfigure}
    \hfill
    \begin{subfigure}[b]{0.11\textwidth}
        \centering
        \includegraphics[width=\textwidth]{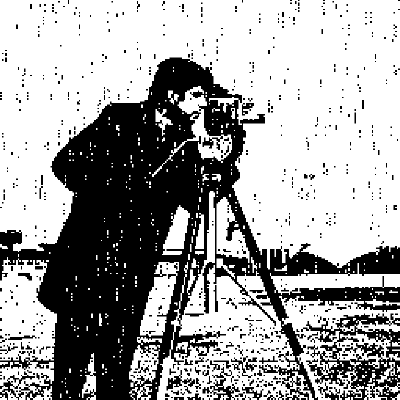}
        %\caption{Noisy image at the receiver of the CIM-IT}
        \caption{}
        \label{case4_noisy}
    \end{subfigure}
    \hfill
    \begin{subfigure}[b]{0.11\textwidth}
        \centering
        \includegraphics[width=\textwidth]{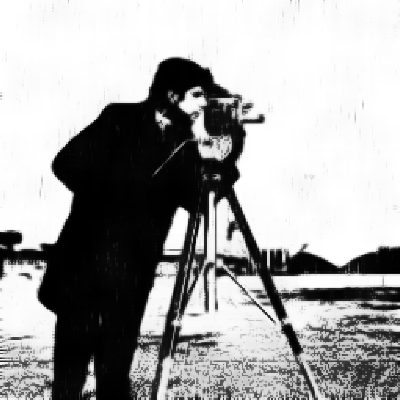}
        %\caption{Image enhanced with NLM denoising technique}
        \caption{}
        \label{case4_enh1}
    \end{subfigure}
    \hfill
    \begin{subfigure}[b]{0.11\textwidth}
        \centering
        \includegraphics[width=\textwidth]{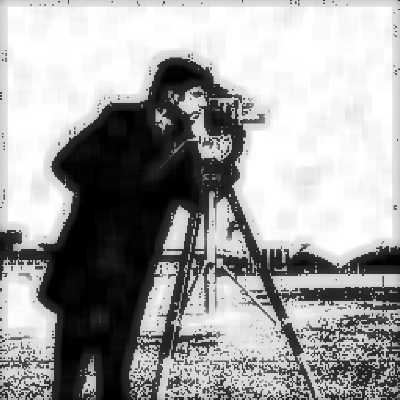}
        %\caption{Image enhanced with Wiener filter}
        \caption{}
        \label{case4_enh2}
    \end{subfigure}
    \hfill
    \begin{subfigure}[b]{0.11\textwidth}
        \centering
        \includegraphics[width=\textwidth]{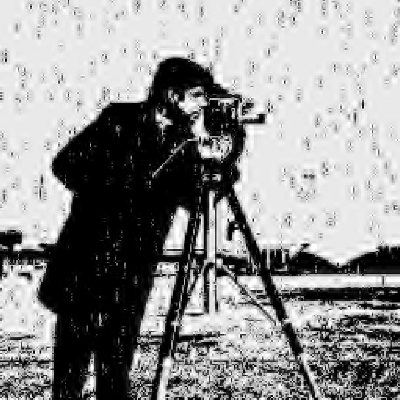}
        %\caption{Image enhanced with wavelet filter}
        \caption{}
        \label{case4_enh3}
    \end{subfigure}
    \hfill
    \begin{subfigure}[b]{0.11\textwidth}
        \centering
        \includegraphics[width=\textwidth]{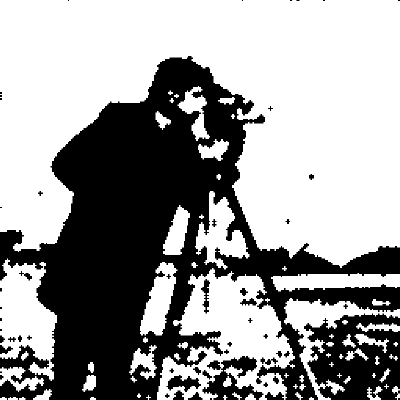}
        %\caption{Image enhanced with morphological filter}
        \caption{}
        \label{case4_enh4}
    \end{subfigure}
    \hfill
    \begin{subfigure}[b]{0.11\textwidth}
        \centering
        \includegraphics[width=\textwidth]{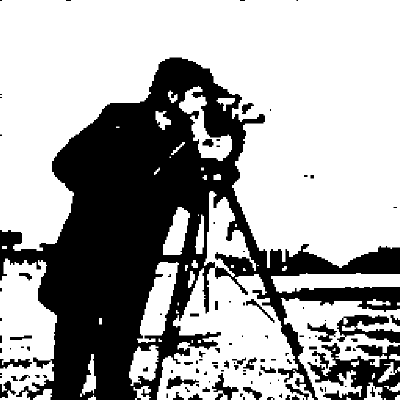}
        %\caption{Image enhanced with majority filter}
        \caption{}
        \label{case4_enh5}
    \end{subfigure}
    \hfill
    \begin{subfigure}[b]{0.11\textwidth}
        \centering
        \includegraphics[width=\textwidth]{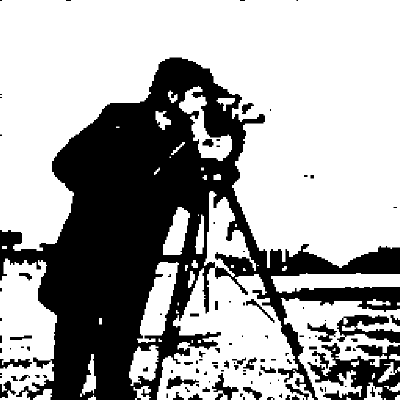}
        %\caption{Image enhanced with median filter}
        \caption{}
        \label{case4_enh6}
    \end{subfigure}
    \caption{Image of the noise at the receiver of the CIM system and images enhanced with different techniques for system parameters  $M=4$, $N_W=3$, $L=16$, $n_r=4$, and $SNR=10$ dB. (a) original image, (b) noisy image at the receiver of the CIM-IT, (c) image enhanced with NLM denoising technique, (d) image enhanced with Wiener filter, (e) image enhanced with wavelet filter, (f) image enhanced with morphological filter, (g) image enhanced with majority filter, and (h) image enhanced with median filter.}
    \label{case4_all}
\end{figure}

Figures \ref{case1_all}-\ref{case4_all} show the comparative analyses of different image enhancement techniques applied to the noisy image received by the proposed CIM-IT system. In the proposed CIM-IT system, the original image is transmitted over the Rayleigh wireless channel and obtained as a noisy image at the receiver. The original image is transmitted by the CIM-IT system under various system parameters and SNR conditions and therefore different noisy images are obtained at the receiver. Fig. \ref{case1_all} and \ref{case2_all} show the performance of the enhancement filters including NLM denoising, Wiener, wavelet, morphological, majority, and median filters applied to noisy images obtained for SNR values of $20$ dB and $10$ dB, respectively, and while system parameters $M=4$, $N_W=3$, $L=16$, $n_r=2$. In Fig. \ref{case1_all}, it can be seen that the noise is quite low because the SNR is high. Therefore, all enhancement techniques except the wavelet effectively remove the noise. The wavelet filter is insufficient to remove the distortions caused by the fading effect of the Rayleigh channel and Gaussian noise. Fig. \ref{case3_all} and Fig. \ref{case4_all} also show the noisy image and its enhanced versions under different SNR conditions and system parameters. Fig. \ref{case3_all} shows the results for an image received with $M=64$, $N_W=5$, $L=128$, $n_r=2$ at an SNR of $20$ dB, while Fig. \ref{case4_all} presents the results for an image received with $M=4$, $N_W=3$, $L=16$, $n_r=4$ at an SNR of $10$ dB. When the results of the enhancement filters are examined, Fig. \ref{case1_all} and Fig. \ref{case2_all}, similar results are obtained.

\section{Conclusions and Discussions}

In conclusion, an index modulation-based image transmission system called CIM-IT utilizing spreading codes and QAM modulation for wireless image transmission has been proposed in this paper. The CIM-IT system maps a bit to each pixel value of the image and transmits them over a wireless channel using a SIMO system. At the receiver, the active spreading code index and selected QAM symbol are estimated, allowing for the reconstruction of the transmitted image. The proposed system has demonstrated several advantages over conventional wireless communication systems. Firstly, it achieves lower bit error rates, leading to improved image transmission quality. Also, throughput, energy efficiency, and spectral efficiency analyses have been conducted, comparing the proposed system with traditional QAM and QPSK systems. The results indicate that the proposed system outperforms these conventional systems in terms of BER for the same spectral efficiency. Furthermore, the performance of the proposed system has been investigated for various system parameters. This analysis helps identify the factors that influence system performance and how they impact the overall results. The proposed system provides potential applications in various fields such as medical and military imaging, where high-resolution, reliable, and high-rate image transmission is crucial. Finally, various enhancement filters are used for the image transmitted by the CIM-IT system through a Rayleigh fading channel and therefore subject to distortions such as fading and noise.


\bibliography{Referanslar.bib}


\begin{thebibliography}{46}
\ifx \bisbn   \undefined \def \bisbn  #1{ISBN #1}\fi
\ifx \binits  \undefined \def \binits#1{#1}\fi
\ifx \bauthor  \undefined \def \bauthor#1{#1}\fi
\ifx \batitle  \undefined \def \batitle#1{#1}\fi
\ifx \bjtitle  \undefined \def \bjtitle#1{#1}\fi
\ifx \bvolume  \undefined \def \bvolume#1{\textbf{#1}}\fi
\ifx \byear  \undefined \def \byear#1{#1}\fi
\ifx \bissue  \undefined \def \bissue#1{#1}\fi
\ifx \bfpage  \undefined \def \bfpage#1{#1}\fi
\ifx \blpage  \undefined \def \blpage #1{#1}\fi
\ifx \burl  \undefined \def \burl#1{\textsf{#1}}\fi
\ifx \doiurl  \undefined \def \doiurl#1{\url{https://doi.org/#1}}\fi
\ifx \betal  \undefined \def \betal{\textit{et al.}}\fi
\ifx \binstitute  \undefined \def \binstitute#1{#1}\fi
\ifx \binstitutionaled  \undefined \def \binstitutionaled#1{#1}\fi
\ifx \bctitle  \undefined \def \bctitle#1{#1}\fi
\ifx \beditor  \undefined \def \beditor#1{#1}\fi
\ifx \bpublisher  \undefined \def \bpublisher#1{#1}\fi
\ifx \bbtitle  \undefined \def \bbtitle#1{#1}\fi
\ifx \bedition  \undefined \def \bedition#1{#1}\fi
\ifx \bseriesno  \undefined \def \bseriesno#1{#1}\fi
\ifx \blocation  \undefined \def \blocation#1{#1}\fi
\ifx \bsertitle  \undefined \def \bsertitle#1{#1}\fi
\ifx \bsnm \undefined \def \bsnm#1{#1}\fi
\ifx \bsuffix \undefined \def \bsuffix#1{#1}\fi
\ifx \bparticle \undefined \def \bparticle#1{#1}\fi
\ifx \barticle \undefined \def \barticle#1{#1}\fi
\bibcommenthead
\ifx \bconfdate \undefined \def \bconfdate #1{#1}\fi
\ifx \botherref \undefined \def \botherref #1{#1}\fi
\ifx \url \undefined \def \url#1{\textsf{#1}}\fi
\ifx \bchapter \undefined \def \bchapter#1{#1}\fi
\ifx \bbook \undefined \def \bbook#1{#1}\fi
\ifx \bcomment \undefined \def \bcomment#1{#1}\fi
\ifx \oauthor \undefined \def \oauthor#1{#1}\fi
\ifx \citeauthoryear \undefined \def \citeauthoryear#1{#1}\fi
\ifx \endbibitem  \undefined \def \endbibitem {}\fi
\ifx \bconflocation  \undefined \def \bconflocation#1{#1}\fi
\ifx \arxivurl  \undefined \def \arxivurl#1{\textsf{#1}}\fi
\csname PreBibitemsHook\endcsname

%%% 1
\bibitem[\protect\citeauthoryear{Tataria et~al.}{2021}]{6GTATARIA}
\begin{barticle}
\bauthor{\bsnm{Tataria}, \binits{H.}},
\bauthor{\bsnm{Shafi}, \binits{M.}},
\bauthor{\bsnm{Molisch}, \binits{A.F.}},
\bauthor{\bsnm{Dohler}, \binits{M.}},
\bauthor{\bsnm{Sjöland}, \binits{H.}},
\bauthor{\bsnm{Tufvesson}, \binits{F.}}:
\batitle{6g wireless systems: Vision, requirements, challenges, insights, and opportunities}.
\bjtitle{Proceedings of the IEEE}
\bvolume{109}(\bissue{7}),
\bfpage{1166}--\blpage{1199}
(\byear{2021})
\doiurl{10.1109/JPROC.2021.3061701}
\end{barticle}
\endbibitem

%%% 2
\bibitem[\protect\citeauthoryear{Bariah et~al.}{2020}]{6GBARIAH}
\begin{barticle}
\bauthor{\bsnm{Bariah}, \binits{L.}},
\bauthor{\bsnm{Mohjazi}, \binits{L.}},
\bauthor{\bsnm{Muhaidat}, \binits{S.}},
\bauthor{\bsnm{Sofotasios}, \binits{P.C.}},
\bauthor{\bsnm{Kurt}, \binits{G.K.}},
\bauthor{\bsnm{Yanikomeroglu}, \binits{H.}},
\bauthor{\bsnm{Dobre}, \binits{O.A.}}:
\batitle{A prospective look: Key enabling technologies, applications and open research topics in 6g networks}.
\bjtitle{IEEE Access}
\bvolume{8},
\bfpage{174792}--\blpage{174820}
(\byear{2020})
\doiurl{10.1109/ACCESS.2020.3019590}
\end{barticle}
\endbibitem

%%% 3
\bibitem[\protect\citeauthoryear{Guo et~al.}{2021}]{6GGUO}
\begin{barticle}
\bauthor{\bsnm{Guo}, \binits{F.}},
\bauthor{\bsnm{Yu}, \binits{F.R.}},
\bauthor{\bsnm{Zhang}, \binits{H.}},
\bauthor{\bsnm{Li}, \binits{X.}},
\bauthor{\bsnm{Ji}, \binits{H.}},
\bauthor{\bsnm{Leung}, \binits{V.C.M.}}:
\batitle{Enabling massive iot toward 6g: A comprehensive survey}.
\bjtitle{IEEE Internet of Things Journal}
\bvolume{8}(\bissue{15}),
\bfpage{11891}--\blpage{11915}
(\byear{2021})
\doiurl{10.1109/JIOT.2021.3063686}
\end{barticle}
\endbibitem

%%% 4
\bibitem[\protect\citeauthoryear{Ozden and Aydin}{2024}]{RISAS}
\begin{botherref}
\oauthor{\bsnm{Ozden}, \binits{B.A.}},
\oauthor{\bsnm{Aydin}, \binits{E.}}:
Antenna selection for receive spatial modulation system empowered by reconfigurable intelligent surface.
IEEE Transactions on Vehicular Technology,
1--11
(2024)
\doiurl{10.1109/TVT.2024.3465573}
\end{botherref}
\endbibitem

%%% 5
\bibitem[\protect\citeauthoryear{Meenalakshmi et~al.}{2022}]{deepmimo}
\begin{bchapter}
\bauthor{\bsnm{Meenalakshmi}, \binits{M.}},
\bauthor{\bsnm{Chaturvedi}, \binits{S.}},
\bauthor{\bsnm{Dwivedi}, \binits{V.K.}}:
\bctitle{Deep learning-based channel estimation in 5g mimo-ofdm systems}.
In: \bbtitle{2022 8th International Conference on Signal Processing and Communication (ICSC)},
pp. \bfpage{79}--\blpage{84}
(\byear{2022}).
\doiurl{10.1109/ICSC56524.2022.10009461}
\end{bchapter}
\endbibitem

%%% 6
\bibitem[\protect\citeauthoryear{Aydın and Ilhan}{2016}]{erdoganSM}
\begin{barticle}
\bauthor{\bsnm{Aydın}, \binits{E.}},
\bauthor{\bsnm{Ilhan}, \binits{H.}}:
\batitle{A novel sm-based mimo system with index modulation}.
\bjtitle{IEEE Communications Letters}
\bvolume{20}(\bissue{2}),
\bfpage{244}--\blpage{247}
(\byear{2016})
\doiurl{10.1109/LCOMM.2015.2512269}
\end{barticle}
\endbibitem

%%% 7
\bibitem[\protect\citeauthoryear{Sundaravadivu and Bharathi}{2013}]{STBCMIMO}
\begin{botherref}
\oauthor{\bsnm{Sundaravadivu}, \binits{K.}},
\oauthor{\bsnm{Bharathi}, \binits{S.}}:
Stbc codes for generalized spatial modulation in mimo systems.
2013 IEEE International Conference ON Emerging Trends in Computing, Communication and Nanotechnology (ICECCN),
486--490
(2013)
\doiurl{10.1109/ICE-CCN.2013.6528548}
\end{botherref}
\endbibitem

%%% 8
\bibitem[\protect\citeauthoryear{Cogen et~al.}{2024}]{fatihanten}
\begin{botherref}
\oauthor{\bsnm{Cogen}, \binits{F.}},
\oauthor{\bsnm{Ozden}, \binits{B.A.}},
\oauthor{\bsnm{Aydin}, \binits{E.}}:
A new energy efficient and antenna selection based index modulation technique.
2024 32nd Signal Processing and Communications Applications Conference (SIU),
1--4
(2024)
\doiurl{10.1109/SIU61531.2024.10601065}
\end{botherref}
\endbibitem

%%% 9
\bibitem[\protect\citeauthoryear{Mao et~al.}{2019}]{IMSURVEY1}
\begin{barticle}
\bauthor{\bsnm{Mao}, \binits{T.}},
\bauthor{\bsnm{Wang}, \binits{Q.}},
\bauthor{\bsnm{Wang}, \binits{Z.}},
\bauthor{\bsnm{Chen}, \binits{S.}}:
\batitle{Novel index modulation techniques: A survey}.
\bjtitle{IEEE Communications Surveys \& Tutorials}
\bvolume{21}(\bissue{1}),
\bfpage{315}--\blpage{348}
(\byear{2019})
\doiurl{10.1109/COMST.2018.2858567}
\end{barticle}
\endbibitem

%%% 10
\bibitem[\protect\citeauthoryear{Doğan~Tusha et~al.}{2021}]{IMSURVEY2}
\begin{barticle}
\bauthor{\bsnm{Doğan~Tusha}, \binits{S.}},
\bauthor{\bsnm{Tusha}, \binits{A.}},
\bauthor{\bsnm{Basar}, \binits{E.}},
\bauthor{\bsnm{Arslan}, \binits{H.}}:
\batitle{Multidimensional index modulation for 5g and beyond wireless networks}.
\bjtitle{Proceedings of the IEEE}
\bvolume{109}(\bissue{2}),
\bfpage{170}--\blpage{199}
(\byear{2021})
\doiurl{10.1109/JPROC.2020.3040589}
\end{barticle}
\endbibitem

%%% 11
\bibitem[\protect\citeauthoryear{Basar}{2016}]{IMSURVEY3}
\begin{barticle}
\bauthor{\bsnm{Basar}, \binits{E.}}:
\batitle{Index modulation techniques for 5g wireless networks}.
\bjtitle{IEEE Communications Magazine}
\bvolume{54}(\bissue{7}),
\bfpage{168}--\blpage{175}
(\byear{2016})
\doiurl{10.1109/MCOM.2016.7509396}
\end{barticle}
\endbibitem

%%% 12
\bibitem[\protect\citeauthoryear{Ozden et~al.}{2023}]{burakris}
\begin{barticle}
\bauthor{\bsnm{Ozden}, \binits{B.A.}},
\bauthor{\bsnm{Aydin}, \binits{E.}},
\bauthor{\bsnm{Cogen}, \binits{F.}}:
\batitle{Reconfigurable intelligent surface-aided spatial media-based modulation}.
\bjtitle{IEEE Transactions on Green Communications and Networking}
\bvolume{7}(\bissue{4}),
\bfpage{1971}--\blpage{1980}
(\byear{2023})
\doiurl{10.1109/TGCN.2023.3289669}
\end{barticle}
\endbibitem

%%% 13
\bibitem[\protect\citeauthoryear{Sugiura et~al.}{2017}]{sugiroindex}
\begin{barticle}
\bauthor{\bsnm{Sugiura}, \binits{S.}},
\bauthor{\bsnm{Ishihara}, \binits{T.}},
\bauthor{\bsnm{Nakao}, \binits{M.}}:
\batitle{State-of-the-art design of index modulation in the space, time, and frequency domains: Benefits and fundamental limitations}.
\bjtitle{IEEE Access}
\bvolume{5},
\bfpage{21774}--\blpage{21790}
(\byear{2017})
\doiurl{10.1109/ACCESS.2017.2763978}
\end{barticle}
\endbibitem

%%% 14
\bibitem[\protect\citeauthoryear{Basar}{2016}]{ertoindex}
\begin{barticle}
\bauthor{\bsnm{Basar}, \binits{E.}}:
\batitle{Index modulation techniques for 5g wireless networks}.
\bjtitle{IEEE Communications Magazine}
\bvolume{54}(\bissue{7}),
\bfpage{168}--\blpage{175}
(\byear{2016})
\doiurl{10.1109/MCOM.2016.7509396}
\end{barticle}
\endbibitem

%%% 15
\bibitem[\protect\citeauthoryear{Li et~al.}{2023}]{whenindex}
\begin{barticle}
\bauthor{\bsnm{Li}, \binits{J.}},
\bauthor{\bsnm{Dang}, \binits{S.}},
\bauthor{\bsnm{Wen}, \binits{M.}},
\bauthor{\bsnm{Li}, \binits{Q.}},
\bauthor{\bsnm{Chen}, \binits{Y.}},
\bauthor{\bsnm{Huang}, \binits{Y.}},
\bauthor{\bsnm{Shang}, \binits{W.}}:
\batitle{Index modulation multiple access for 6g communications: Principles, applications, and challenges}.
\bjtitle{IEEE Network}
\bvolume{37}(\bissue{1}),
\bfpage{52}--\blpage{60}
(\byear{2023})
\doiurl{10.1109/MNET.002.2200433}
\end{barticle}
\endbibitem

%%% 16
\bibitem[\protect\citeauthoryear{Ozden et~al.}{2023}]{codeozden}
\begin{barticle}
\bauthor{\bsnm{Ozden}, \binits{B.A.}},
\bauthor{\bsnm{Cogen}, \binits{F.}},
\bauthor{\bsnm{Aydin}, \binits{E.}}:
\batitle{Mirror activation pattern selection for energy efficient hexagonal qam aided media-based modulation}.
\bjtitle{Transactions on Emerging Telecommunications Technologies}
\bvolume{34}(\bissue{7}),
\bfpage{4795}
(\byear{2023})
\doiurl{10.1002/ett.4795}
{\href{https://arxiv.org/abs/https://onlinelibrary.wiley.com/doi/pdf/10.1002/ett.4795}{{https://onlinelibrary.wiley.com/doi/pdf/10.1002/ett.4795}}}
\end{barticle}
\endbibitem

%%% 17
\bibitem[\protect\citeauthoryear{Frank et~al.}{2007}]{CDMA1}
\begin{barticle}
\bauthor{\bsnm{Frank}, \binits{T.}},
\bauthor{\bsnm{Klein}, \binits{A.}},
\bauthor{\bsnm{Costa}, \binits{E.}}:
\batitle{Ifdma: A scheme combining the advantages of ofdma and cdma}.
\bjtitle{IEEE Wireless Communications}
\bvolume{14}(\bissue{3}),
\bfpage{9}--\blpage{17}
(\byear{2007})
\doiurl{10.1109/MWC.2007.386607}
\end{barticle}
\endbibitem

%%% 18
\bibitem[\protect\citeauthoryear{Baier}{1996}]{CDMA2}
\begin{barticle}
\bauthor{\bsnm{Baier}, \binits{P.W.}}:
\batitle{A critical review of cdma}.
\bjtitle{Proceedings of Vehicular Technology Conference - VTC}
\bvolume{1},
\bfpage{6}--\blpage{101}
(\byear{1996})
\doiurl{10.1109/VETEC.1996.503397}
\end{barticle}
\endbibitem

%%% 19
\bibitem[\protect\citeauthoryear{Gilhousen et~al.}{1991}]{CDMA3}
\begin{barticle}
\bauthor{\bsnm{Gilhousen}, \binits{K.S.}},
\bauthor{\bsnm{Jacobs}, \binits{I.M.}},
\bauthor{\bsnm{Padovani}, \binits{R.}},
\bauthor{\bsnm{Viterbi}, \binits{A.J.}},
\bauthor{\bsnm{Weaver}, \binits{L.A.}},
\bauthor{\bsnm{Wheatley}, \binits{C.E.}}:
\batitle{On the capacity of a cellular cdma system}.
\bjtitle{IEEE Transactions on Vehicular Technology}
\bvolume{40}(\bissue{2}),
\bfpage{303}--\blpage{312}
(\byear{1991})
\doiurl{10.1109/25.289411}
\end{barticle}
\endbibitem

%%% 20
\bibitem[\protect\citeauthoryear{Kase et~al.}{2017}]{CDMA4}
\begin{botherref}
\oauthor{\bsnm{Kase}, \binits{Y.}},
\oauthor{\bsnm{Tanaka}, \binits{M.}},
\oauthor{\bsnm{Seki}, \binits{T.}}:
Study on characteristics of mc-cdma communication system.
2017 IEEE Asia Pacific Microwave Conference (APMC),
1018--1021
(2017)
\doiurl{10.1109/APMC.2017.8251625}
\end{botherref}
\endbibitem

%%% 21
\bibitem[\protect\citeauthoryear{Wang et~al.}{2018}]{CDMA5}
\begin{barticle}
\bauthor{\bsnm{Wang}, \binits{J.}},
\bauthor{\bsnm{Guo}, \binits{S.}},
\bauthor{\bsnm{Chen}, \binits{Z.}},
\bauthor{\bsnm{Li}, \binits{Y.}},
\bauthor{\bsnm{Lu}, \binits{Z.}}:
\batitle{A new parallel codec technique for cdma nocs}.
\bjtitle{IEEE Transactions on Industrial Electronics}
\bvolume{65}(\bissue{8}),
\bfpage{6527}--\blpage{6537}
(\byear{2018})
\doiurl{10.1109/TIE.2017.2786230}
\end{barticle}
\endbibitem

%%% 22
\bibitem[\protect\citeauthoryear{Cogen et~al.}{2018}]{Cogen2018siu}
\begin{botherref}
\oauthor{\bsnm{Cogen}, \binits{F.}},
\oauthor{\bsnm{Aydin}, \binits{E.}},
\oauthor{\bsnm{Kabaoglu}, \binits{N.}},
\oauthor{\bsnm{Başar}, \binits{E.}},
\oauthor{\bsnm{Ilhan}, \binits{H.}}:
{A novel MIMO scheme based on code-index modulation and spatial modulation}.
26th IEEE Signal Processing and Communications Applications Conference, SIU 2018,
1--4
(2018)
\end{botherref}
\endbibitem

%%% 23
\bibitem[\protect\citeauthoryear{Ozden et~al.}{2024}]{burakwcnc}
\begin{botherref}
\oauthor{\bsnm{Ozden}, \binits{B.A.}},
\oauthor{\bsnm{Cogen}, \binits{F.}},
\oauthor{\bsnm{Aydin}, \binits{E.}},
\oauthor{\bsnm{Ilhan}, \binits{H.}},
\oauthor{\bsnm{Basar}, \binits{E.}},
\oauthor{\bsnm{Wen}, \binits{M.}}:
A novel reconfigurable intelligent surface-supported code index modulation-based receive spatial modulation system.
2024 IEEE Wireless Communications and Networking Conference (WCNC),
1--6
(2024)
\doiurl{10.1109/WCNC57260.2024.10571099}
\end{botherref}
\endbibitem

%%% 24
\bibitem[\protect\citeauthoryear{Zhai et~al.}{2021}]{6gcim}
\begin{botherref}
\oauthor{\bsnm{Zhai}, \binits{W.}},
\oauthor{\bsnm{Wu}, \binits{Y.}},
\oauthor{\bsnm{Zhao}, \binits{J.}},
\oauthor{\bsnm{Han}, \binits{H.}}:
6g downlink transmission via rate splitting space division multiple access based on grouped code index modulation.
2021 IEEE International Conference on Communications Workshops (ICC Workshops),
1--6
(2021)
\doiurl{10.1109/ICCWorkshops50388.2021.9473603}
\end{botherref}
\endbibitem

%%% 25
\bibitem[\protect\citeauthoryear{Kaddoum et~al.}{2016}]{kadoumcim}
\begin{barticle}
\bauthor{\bsnm{Kaddoum}, \binits{G.}},
\bauthor{\bsnm{Nijsure}, \binits{Y.}},
\bauthor{\bsnm{Tran}, \binits{H.}}:
\batitle{Generalized code index modulation technique for high-data-rate communication systems}.
\bjtitle{IEEE Transactions on Vehicular Technology}
\bvolume{65}(\bissue{9}),
\bfpage{7000}--\blpage{7009}
(\byear{2016})
\doiurl{10.1109/TVT.2015.2498040}
\end{barticle}
\endbibitem

%%% 26
\bibitem[\protect\citeauthoryear{Cogen et~al.}{2024}]{fatihris}
\begin{barticle}
\bauthor{\bsnm{Cogen}, \binits{F.}},
\bauthor{\bsnm{Ozden}, \binits{B.A.}},
\bauthor{\bsnm{Aydin}, \binits{E.}},
\bauthor{\bsnm{Kabaoglu}, \binits{N.}}:
\batitle{Reconfigurable intelligent surface-empowered code index modulation for high-rate siso systems}.
\bjtitle{IEEE Transactions on Cognitive Communications and Networking}
\bvolume{10}(\bissue{5}),
\bfpage{1856}--\blpage{1866}
(\byear{2024})
\doiurl{10.1109/TCCN.2024.3384495}
\end{barticle}
\endbibitem

%%% 27
\bibitem[\protect\citeauthoryear{Ozden et~al.}{2023}]{burakcim}
\begin{barticle}
\bauthor{\bsnm{Ozden}, \binits{B.A.}},
\bauthor{\bsnm{Aydin}, \binits{E.}},
\bauthor{\bsnm{Cogen}, \binits{F.}}:
\batitle{Code index modulation-aided spatial media-based modulation system for future wireless networks}.
\bjtitle{IEEE Systems Journal}
\bvolume{17}(\bissue{3}),
\bfpage{3762}--\blpage{3770}
(\byear{2023})
\doiurl{10.1109/JSYST.2022.3230775}
\end{barticle}
\endbibitem

%%% 28
\bibitem[\protect\citeauthoryear{Xu et~al.}{2017}]{CIMGCSDCSK}
\begin{barticle}
\bauthor{\bsnm{Xu}, \binits{W.}},
\bauthor{\bsnm{Huang}, \binits{T.}},
\bauthor{\bsnm{Wang}, \binits{L.}}:
\batitle{Code-shifted differential chaos shift keying with code index modulation for high data rate transmission}.
\bjtitle{IEEE Transactions on Communications}
\bvolume{65}(\bissue{10}),
\bfpage{4285}--\blpage{4294}
(\byear{2017})
\doiurl{10.1109/TCOMM.2017.2725261}
\end{barticle}
\endbibitem

%%% 29
\bibitem[\protect\citeauthoryear{Aydin et~al.}{2019}]{aydinQSM}
\begin{barticle}
\bauthor{\bsnm{Aydin}, \binits{E.}},
\bauthor{\bsnm{Cogen}, \binits{F.}},
\bauthor{\bsnm{Basar}, \binits{E.}}:
\batitle{Code-index modulation aided quadrature spatial modulation for high-rate mimo systems}.
\bjtitle{IEEE Transactions on Vehicular Technology}
\bvolume{68}(\bissue{10}),
\bfpage{10257}--\blpage{10261}
(\byear{2019})
\doiurl{10.1109/TVT.2019.2928378}
\end{barticle}
\endbibitem

%%% 30
\bibitem[\protect\citeauthoryear{Bourtsoulatze et~al.}{2019}]{imagesurvey1}
\begin{botherref}
\oauthor{\bsnm{Bourtsoulatze}, \binits{E.}},
\oauthor{\bsnm{Kurka}, \binits{D.B.}},
\oauthor{\bsnm{Gündüz}, \binits{D.}}:
Deep joint source-channel coding for wireless image transmission.
ICASSP 2019 - 2019 IEEE International Conference on Acoustics, Speech and Signal Processing (ICASSP),
4774--4778
(2019)
\doiurl{10.1109/ICASSP.2019.8683463}
\end{botherref}
\endbibitem

%%% 31
\bibitem[\protect\citeauthoryear{Tung and Gündüz}{2018}]{imagesurvey2}
\begin{barticle}
\bauthor{\bsnm{Tung}, \binits{T.-Y.}},
\bauthor{\bsnm{Gündüz}, \binits{D.}}:
\batitle{Sparsecast: Hybrid digital-analog wireless image transmission exploiting frequency-domain sparsity}.
\bjtitle{IEEE Communications Letters}
\bvolume{22}(\bissue{12}),
\bfpage{2451}--\blpage{2454}
(\byear{2018})
\doiurl{10.1109/LCOMM.2018.2877316}
\end{barticle}
\endbibitem

%%% 32
\bibitem[\protect\citeauthoryear{Al-Hayani and Ilhan}{2020}]{haciimage}
\begin{barticle}
\bauthor{\bsnm{Al-Hayani}, \binits{B.}},
\bauthor{\bsnm{Ilhan}, \binits{H.}}:
\batitle{Efficient cooperative image transmission in one-way multi-hop sensor network}.
\bjtitle{International Journal of Electrical Engineering \& Education}
\bvolume{57}(\bissue{4}),
\bfpage{321}--\blpage{339}
(\byear{2020})
\doiurl{10.1177/0020720918816009}
\end{barticle}
\endbibitem

%%% 33
\bibitem[\protect\citeauthoryear{Ou et~al.}{2020}]{zhengimage}
\begin{barticle}
\bauthor{\bsnm{Ou}, \binits{L.}},
\bauthor{\bsnm{Liao}, \binits{S.}},
\bauthor{\bsnm{Qin}, \binits{Z.}},
\bauthor{\bsnm{Yin}, \binits{H.}}:
\batitle{Millimeter wave wireless hadamard image transmission for mimo enabled 5g and beyond}.
\bjtitle{IEEE Wireless Communications}
\bvolume{27}(\bissue{6}),
\bfpage{134}--\blpage{139}
(\byear{2020})
\doiurl{10.1109/MWC.001.2000081}
\end{barticle}
\endbibitem

%%% 34
\bibitem[\protect\citeauthoryear{Xu et~al.}{2022}]{imaiterat1}
\begin{barticle}
\bauthor{\bsnm{Xu}, \binits{J.}},
\bauthor{\bsnm{Ai}, \binits{B.}},
\bauthor{\bsnm{Chen}, \binits{W.}},
\bauthor{\bsnm{Yang}, \binits{A.}},
\bauthor{\bsnm{Sun}, \binits{P.}},
\bauthor{\bsnm{Rodrigues}, \binits{M.}}:
\batitle{Wireless image transmission using deep source channel coding with attention modules}.
\bjtitle{IEEE Transactions on Circuits and Systems for Video Technology}
\bvolume{32}(\bissue{4}),
\bfpage{2315}--\blpage{2328}
(\byear{2022})
\doiurl{10.1109/TCSVT.2021.3082521}
\end{barticle}
\endbibitem

%%% 35
\bibitem[\protect\citeauthoryear{Ye et~al.}{2021}]{imaiterat2}
\begin{barticle}
\bauthor{\bsnm{Ye}, \binits{H.}},
\bauthor{\bsnm{Li}, \binits{G.Y.}},
\bauthor{\bsnm{Juang}, \binits{B.-H.}}:
\batitle{Deep learning based end-to-end wireless communication systems without pilots}.
\bjtitle{IEEE Transactions on Cognitive Communications and Networking}
\bvolume{7}(\bissue{3}),
\bfpage{702}--\blpage{714}
(\byear{2021})
\doiurl{10.1109/TCCN.2021.3061464}
\end{barticle}
\endbibitem

%%% 36
\bibitem[\protect\citeauthoryear{Baharuddin and Angraini}{2019}]{imaiterat3}
\begin{barticle}
\bauthor{\bsnm{Baharuddin}},
\bauthor{\bsnm{Angraini}, \binits{R.}}:
\batitle{Performance evaluation of image transmission using diversity selection combining technique}.
\bjtitle{IOP Conference Series: Materials Science and Engineering}
\bvolume{602}(\bissue{1}),
\bfpage{012014}
(\byear{2019})
\doiurl{10.1088/1757-899X/602/1/012014}
\end{barticle}
\endbibitem

%%% 37
\bibitem[\protect\citeauthoryear{Song et~al.}{2017}]{imaiterat4}
\begin{barticle}
\bauthor{\bsnm{Song}, \binits{X.}},
\bauthor{\bsnm{Peng}, \binits{X.}},
\bauthor{\bsnm{Xu}, \binits{J.}},
\bauthor{\bsnm{Shi}, \binits{G.}},
\bauthor{\bsnm{Wu}, \binits{F.}}:
\batitle{Distributed compressive sensing for cloud-based wireless image transmission}.
\bjtitle{IEEE Transactions on Multimedia}
\bvolume{19}(\bissue{6}),
\bfpage{1351}--\blpage{1364}
(\byear{2017})
\doiurl{10.1109/TMM.2017.2654123}
\end{barticle}
\endbibitem

%%% 38
\bibitem[\protect\citeauthoryear{Singh et~al.}{2021}]{imaiterat5}
\begin{barticle}
\bauthor{\bsnm{Singh}, \binits{M.}},
\bauthor{\bsnm{Singh}, \binits{M.L.}},
\bauthor{\bsnm{Singh}, \binits{G.}},
\bauthor{\bsnm{Kaur}, \binits{H.}},
\bauthor{\bsnm{Priyanka}},
\bauthor{\bsnm{Kaur}, \binits{S.}}:
\batitle{Real-time image transmission through underwater wireless optical communication link for internet of underwater things}.
\bjtitle{International Journal of Communication Systems}
\bvolume{34}(\bissue{16}),
\bfpage{4951}
(\byear{2021})
\doiurl{10.1002/dac.4951}
{\href{https://arxiv.org/abs/https://onlinelibrary.wiley.com/doi/pdf/10.1002/dac.4951}{{https://onlinelibrary.wiley.com/doi/pdf/10.1002/dac.4951}}}
\end{barticle}
\endbibitem

%%% 39
\bibitem[\protect\citeauthoryear{Abdollahi et~al.}{2021}]{sonimage1}
\begin{barticle}
\bauthor{\bsnm{Abdollahi}, \binits{N.}},
\bauthor{\bsnm{Shahtalebi}, \binits{K.}},
\bauthor{\bsnm{Sabahi}, \binits{M.F.}}:
\batitle{High compression rate, based on the rls adaptive algorithm in progressive image transmission}.
\bjtitle{Signal, Image and Video Processing}
\bvolume{15}(\bissue{4}),
\bfpage{835}--\blpage{842}
(\byear{2021})
\end{barticle}
\endbibitem

%%% 40
\bibitem[\protect\citeauthoryear{Yang et~al.}{2022}]{sonimage2}
\begin{barticle}
\bauthor{\bsnm{Yang}, \binits{H.}},
\bauthor{\bsnm{Qing}, \binits{L.}},
\bauthor{\bsnm{Yang}, \binits{J.}},
\bauthor{\bsnm{He}, \binits{X.}}:
\batitle{Progressively refined scheme for wireless video sensor networks}.
\bjtitle{Signal, Image and Video Processing}
\bvolume{16}(\bissue{6}),
\bfpage{1435}--\blpage{1442}
(\byear{2022})
\end{barticle}
\endbibitem

%%% 41
\bibitem[\protect\citeauthoryear{Tse and Viswanath}{2005}]{Tse}
\begin{bbook}
\bauthor{\bsnm{Tse}, \binits{D.}},
\bauthor{\bsnm{Viswanath}, \binits{P.}}:
\bbtitle{Fundamentals of Wireless Communication}.
\bpublisher{Cambridge University Press}, \blocation{???}
(\byear{2005}).
\doiurl{10.1017/CBO9780511807213}
\end{bbook}
\endbibitem

%%% 42
\bibitem[\protect\citeauthoryear{Gonzalez}{2009}]{Gonzalez2002}
\begin{bbook}
\bauthor{\bsnm{Gonzalez}, \binits{R.C.}}:
\bbtitle{Digital Image Processing}.
\bpublisher{Pearson education india}, \blocation{???}
(\byear{2009})
\end{bbook}
\endbibitem

%%% 43
\bibitem[\protect\citeauthoryear{Buades et~al.}{2005}]{Buades2005}
\begin{barticle}
\bauthor{\bsnm{Buades}, \binits{A.}},
\bauthor{\bsnm{Coll}, \binits{B.}},
\bauthor{\bsnm{Morel}, \binits{J.-M.}}:
\batitle{A non-local algorithm for image denoising}.
\bjtitle{2005 IEEE Computer Society Conference on Computer Vision and Pattern Recognition (CVPR'05)}
\bvolume{2},
\bfpage{60}--\blpage{652}
(\byear{2005})
\doiurl{10.1109/CVPR.2005.38}
\end{barticle}
\endbibitem

%%% 44
\bibitem[\protect\citeauthoryear{Lee}{1980}]{Lee1980}
\begin{barticle}
\bauthor{\bsnm{Lee}, \binits{J.-S.}}:
\batitle{Digital image enhancement and noise filtering by use of local statistics}.
\bjtitle{IEEE Transactions on Pattern Analysis and Machine Intelligence}
\bvolume{PAMI-2}(\bissue{2}),
\bfpage{165}--\blpage{168}
(\byear{1980})
\doiurl{10.1109/TPAMI.1980.4766994}
\end{barticle}
\endbibitem

%%% 45
\bibitem[\protect\citeauthoryear{Mallat}{1989}]{Mallat1989}
\begin{barticle}
\bauthor{\bsnm{Mallat}, \binits{S.G.}}:
\batitle{A theory for multiresolution signal decomposition: the wavelet representation}.
\bjtitle{IEEE Transactions on Pattern Analysis and Machine Intelligence}
\bvolume{11}(\bissue{7}),
\bfpage{674}--\blpage{693}
(\byear{1989})
\doiurl{10.1109/34.192463}
\end{barticle}
\endbibitem

%%% 46
\bibitem[\protect\citeauthoryear{Serra}{1983}]{Serra1982}
\begin{bbook}
\bauthor{\bsnm{Serra}, \binits{J.}}:
\bbtitle{Image Analysis and Mathematical Morphology}.
\bpublisher{Academic Press, Inc.}, \blocation{???}
(\byear{1983})
\end{bbook}
\endbibitem

\end{thebibliography}

  \begin{wrapfigure}{l}{22mm} 
\includegraphics[width=1.5in,height=1.5in,clip,keepaspectratio]{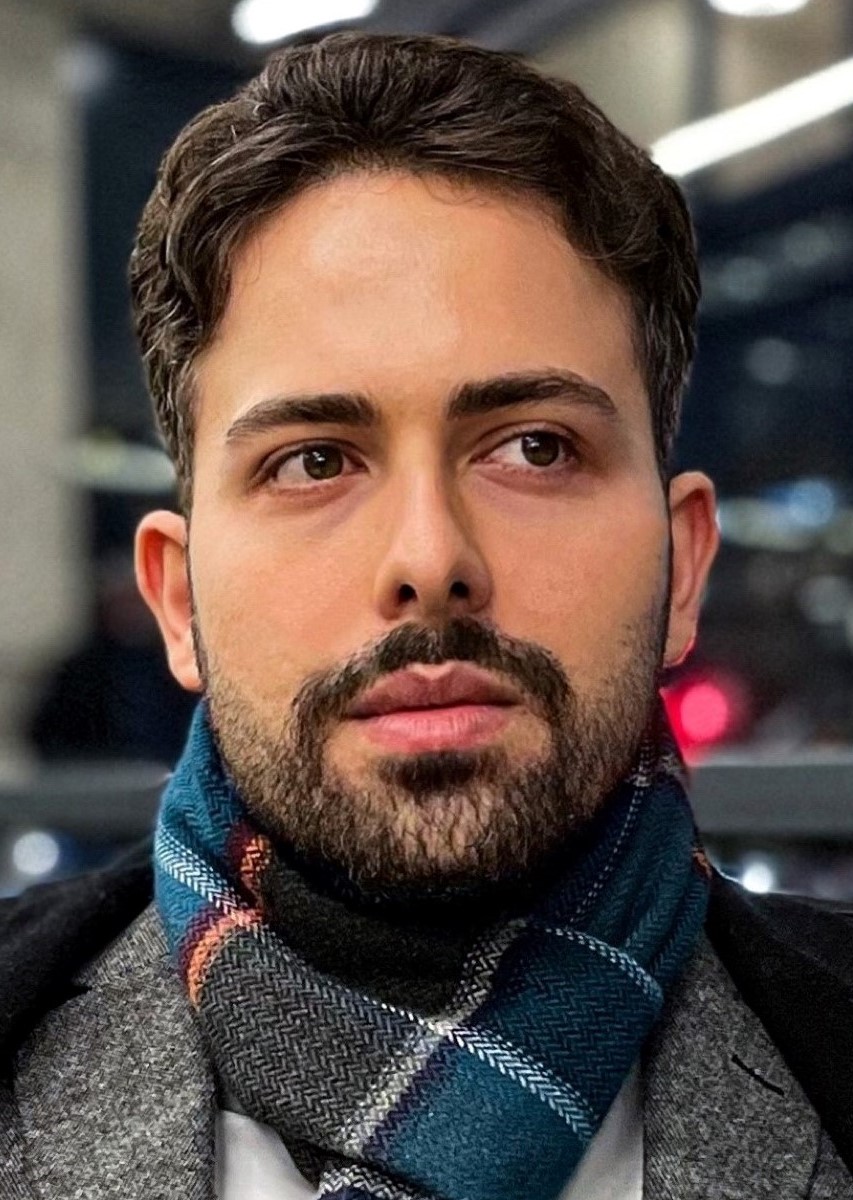}
\end{wrapfigure}\par
\textbf{Burak Ahmet Ozden} was born in Istanbul, Turkey. He received a B.S. with high honors and an M.S. degree in Electrical and Electronic Engineering from Istanbul Medeniyet University in 2020 and 2022, respectively. He is currently pursuing a Ph.D. in Electrical and Electronic Engineering at Istanbul Medeniyet University while also working as a research assistant at Yıldız Teknik University. His research interests comprise index modulation, wireless communication, reconfigurable intelligent surfaces, signal processing, MIMO systems, and cooperative communication. Also, Burak Ahmet Ozden serves as a reviewer for many journals, such as IEEE Transactions on Wireless Communications, IEEE Internet of Things Journal, IEEE Transactions on Vehicular Technology, and IEEE Wireless Communications Letters.  \par

\begin{wrapfigure}{l}{22mm} 
    \includegraphics[width=1.5in,height=1.5in,clip,keepaspectratio]{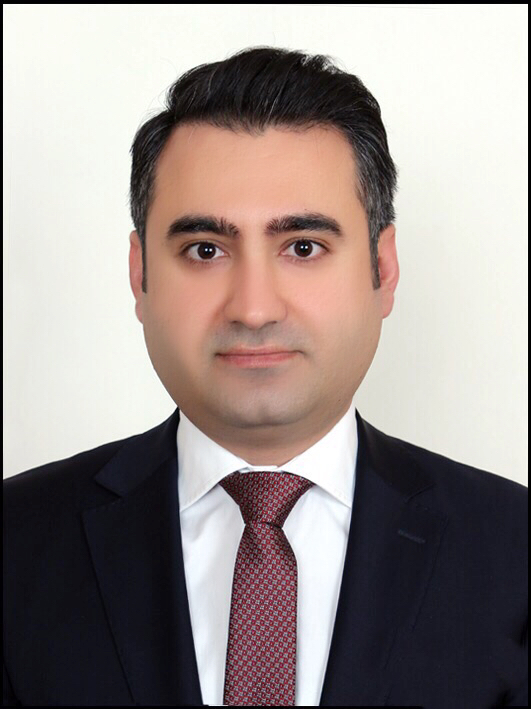}
  \end{wrapfigure}\par
  \textbf{Erdogan Aydin} was born in Turkey. He received the B.S. with high honors and the M.S. degrees from Istanbul University, Istanbul, Turkey, in 2007 and 2010, respectively, and the Ph.D. degree from Yıldız Technical University, Istanbul, Turkey, in 2016. He is currently an Associate Professor  with the Department of Electronics and Communication Engineering, Istanbul Medeniyet University, Istanbul, Turkey. His primary research interests include  MIMO systems, cooperative communication and diversity, index modulation, spatial modulation, media based modulation, hexagonal constellations, non-orthogonal multiple access, visible light communication, chaos communication, statistical signal processing, and estimation theory. He has received best paper award including one from the IEEE International Conference on Communications 2018. He has served as a TPC member for several IEEE conferences. \par

\begin{wrapfigure}{l}{22mm} 
\includegraphics[width=1.in,height=1.9in,clip,keepaspectratio]{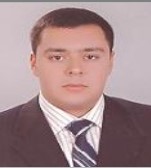}
\end{wrapfigure}\par
\textbf{Ahmet Elbir}  was born in Konya, Turkey. He received the B.S. degree in Computer and Control Teaching from Marmara University in 2012, and the M.S. and Ph.D. degrees in Computer Engineering from Yıldız Technical University in 2015 and 2020, respectively. He is currently a Lecturer in the Department of Computer Engineering at Yıldız Technical University. His research interests include operating systems, data mining, algorithms and programming, music signal processing, and machine learning. \par

    \begin{wrapfigure}{l}{22mm} 
    \includegraphics[width=1.5in,height=1.5in,clip,keepaspectratio]{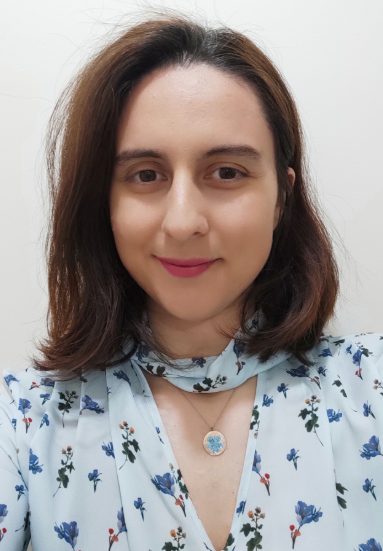}
  \end{wrapfigure}\par
  \textbf{Filiz Gurkan}  received PhD degree in Electronics and Communication Engineering from Istanbul Technical University, Istanbul, Turkey in 2021. Since 2022 she has been working as an assistant professor at Medeniyet University, Turkey. Her research interests include image processing, pattern recognition, machine learning,  particularly in deep learning for object tracking and classification. \par

%\input sn-article.bbl

%\input{sn-article.bbl}
\end{document}